\documentclass[twocolumn,preprintnumbers,a4paper,
superscriptaddress,floatfix,twoside]{revtex4}

\usepackage{bm}
\usepackage{epsfig}
\usepackage{graphics}
\usepackage{natbib}
\usepackage{longtable}
\usepackage{fancyh}
\usepackage[l]{floatflt}
\usepackage{epsfig}
\usepackage{amssymb}
\usepackage{latexsym}
\usepackage{times}
\usepackage{slashed}
\usepackage{upgreek}
\usepackage{amsmath}
\usepackage{amsfonts}
\usepackage{amsbsy}
\usepackage{amscd}
\usepackage{bbm}

\newcommand{\Eref}[1]{Eq.~(\ref{#1})}
\newcommand{\Erefs}[2]{Eqs.~(\ref{#1}) -- (\ref{#2})}
\newcommand{\Sref}[1]{Sec.~\ref{#1}}
\newcommand{\Srefs}[1]{Secs.~\ref{#1}}
\newcommand{\Fref}[1]{Fig.~\ref{#1}}
\newcommand{\Frefs}[1]{Figs.~\ref{#1}}
\newcommand{\Tref}[1]{Table~\ref{#1}}
\newcommand{\Trefs}[1]{Table~\ref{#1}}
\newcommand{\cref}[1]{Ref.~\cite{#1}}
\newcommand{\crefs}[1]{Refs.~\cite{#1}}

\newcommand{\hepth}[1]{{\ftn \tt hep-th/#1}}
\newcommand{\hepph}[1]{{\ftn\tt hep-ph/#1}}

\newcommand{\astroph}[1]{{\ftn\tt astro-ph/#1}}
\newcommand{\arxiv}[1]{{\ftn\tt  arXiv:#1}}

\newcommand{\bal}{\begin{align}}
\newcommand{\eal}{\end{align}}
\newcommand{\beqs}{\begin{subequations}}
\newcommand{\eeqs}{\end{subequations}}
\newcommand{\eec}{\end{center}}
\newcommand{\bec}{\begin{center}}

\newcommand{\eem}{\end{matrix}}
\newcommand{\bem}{\begin{matrix}}
\newcommand{\eeq}{\end{equation}}
\newcommand{\beq}{\begin{equation}}
\newcommand{\ba}{\begin{array}}
\newcommand{\ea}{\end{array}}
\newcommand{\bea}{\begin{eqnarray}}
\newcommand{\eea}{\end{eqnarray}}
\newcommand{\baq}{\begin{eqnarray}}
\newcommand{\eaq}{\end{eqnarray}}

\newcommand\eqs[2]{Eqs.~(\ref{#1}) and (\ref{#2})}

\newcommand\eqss[3]{Eqs.~(\ref{#1}), (\ref{#2}) and (\ref{#3})}

\newcommand{\ftn}{\footnotesize}

\newcommand{\TeV}{{\mbox{\rm TeV}}}

\newcommand{\GeV}{{\mbox{\rm GeV}}}
\newcommand{\eV}{{\mbox{\rm eV}}}

\newcommand{\sFref}[2]{Fig.~\ref{#1}-{\ftn\sf ({#2})}}

\newcommand{\etal}{{\it et al.\/}}

\def\to{\rightarrow}

\def\llgm{\left\lgroup}
\def\rrgm{\right\rgroup}
\def\lf{\left(}
\def\rg{\right)}
\newcommand\vev[1]{\langle {#1} \rangle}
\newcommand{\Gr}{\ensuremath{\widetilde{G}}}
\newcommand{\Yb}{\ensuremath{Y_{B}}}
\newcommand{\Yg}{\ensuremath{Y_{3/2}}}
\newcommand{\Vhi}{\ensuremath{\widehat V_{\rm HI}}}

\newcommand{\Hhi}{\ensuremath{\widehat H_{\rm HI}}}
\newcommand{\Ohi}{\ensuremath{\Omega}}
\newcommand{\Omg}{\ensuremath{\Omega}}
\newcommand{\Khi}{\ensuremath{K}}

\newcommand{\kar}{\ensuremath{K_{1\cal R}}}
\newcommand{\kbr}{\ensuremath{K_{2\cal R}}}

\newcommand{\Ns}{\ensuremath{{\what N_\star}}}

\newcommand{\mP}{\ensuremath{m_{\rm P}}}

\newcommand{\Mgut}{\ensuremath{M_{\rm GUT}}}
\newcommand{\Qef}{\ensuremath{\Lambda_{\rm UV}}}

\newcommand{\lm}{\ensuremath{\lambda_\mu}}

\def\openone{\leavevmode\hbox{\small1\kern-3.8pt\normalsize1}}
\newcommand{\dV}{\ensuremath{\Delta\widehat V_{\rm HI}}}
\newcommand{\Gbl}{\ensuremath{G_{B-L}}}
\newcommand{\bl}{\ensuremath{U(1)_{B-L}}}

\newcommand{\fw}{\ensuremath{f_{\rm W}}}
\newcommand{\tfw}{\ensuremath{\tilde f_{\rm W}}}
\newcommand{\fr}{\ensuremath{f_{\cal R}}}
\newcommand{\frs}{\ensuremath{f_{\cal R\star}}}

\newcommand{\fk}{\ensuremath{f_{\rm K}}}
\newcommand{\fm}{\ensuremath{F_{-}}}
\newcommand{\fp}{\ensuremath{F_{+}}}
\newcommand{\hr}{\ensuremath{F_{\cal R}}}

\newcommand{\ca}{\ensuremath{c_{\cal R}}}

\newcommand{\Gsn}{\ensuremath{\what{\Gamma}_{\rm \dph}}}
\newcommand{\GNsn}{\ensuremath{\what{\Gamma}_{\dph\to N_i^cN_i^c}}}
\newcommand{\Ghsn}{\ensuremath{\what{\Gamma}_{\dph\to\hu\hd}}}
\newcommand{\Gysn}{\ensuremath{\what{\Gamma}_{\dph\to XYZ}}}
\newcommand{\msn}{\ensuremath{\what m_{\rm \dph}}}
\newcommand{\aS}{\ensuremath{{\rm a}_S}}
\newcommand{\Ald}{\ensuremath{A_\lambda}}

\newcommand{\hd}{{\ensuremath{H_d}}}
\newcommand{\hu}{{\ensuremath{H_u}}}

\newcommand{\ks}{\ensuremath{k_\star}}
\newcommand{\ns}{\ensuremath{n_{\rm s}}}

\newcommand{\as}{\ensuremath{a_{\rm s}}}
\newcommand{\As}{\ensuremath{A_{\rm s}}}
\newcommand{\rw}{\ensuremath{r_{0.002}}}
\newcommand{\rs}{\ensuremath{r_{\pm}}}
\newcommand{\rsb}{\ensuremath{\bar r_{\pm}}}

\newcommand{\rcc}{\ensuremath{\mathcal{R}}}
\newcommand{\rce}{\ensuremath{\widehat{\mathcal{R}}}}
\newcommand{\Ve}{\ensuremath{\widehat{V}}}

\newcommand{\sni}{\ensuremath{N^c_i}}
\newcommand{\ssni}{\ensuremath{\widetilde N^c_i}}

\newcommand{\Dex}{\ensuremath{\Delta_{\rm max\star}}}

\newcommand{\mrh[1]}{\ensuremath{M_{#1N^c}}}
\newcommand{\lrh[1]}{\ensuremath{\lambda_{#1N^c}}}

\newcommand{\mD[1]}{\ensuremath{m_{#1\rm D}}}
\newcommand{\mn[1]}{\ensuremath{m_{#1\rm \nu}}}

\newcommand{\wrhn[1]}{\ensuremath{N^c_{#1}}}
\newcommand{\nsu}{\ensuremath{{N_X}}}

\newcommand{\dphi}{\ensuremath{\what{\delta\phi}}}
\newcommand{\dph}{\ensuremath{\delta\phi}}

\newcommand{\phc}{\ensuremath{\Phi}}
\newcommand{\phcb}{\ensuremath{\bar\Phi}}

\newcommand{\what}{\ensuremath{\widehat}}
\newcommand{\wtilde}{\ensuremath{\widetilde}}

\newcommand{\Wmu}{\ensuremath{W_{\mu}}}
\newcommand{\Wrhn}{\ensuremath{W_{\rm RHN}}}
\newcommand{\Whi}{\ensuremath{W_{\rm HI}}}
\def\ve{\varepsilon}
\def\aal{{\bar\alpha}}
\def\bbet{{\bar\beta}}
\def\al{{\alpha}}

\def\bt{{\beta}}

\def\n{\bar{n}}

\def\th{{\theta}}
\def\thb{{\bar\theta}}

\def\thn{{\theta_{\Phi}}}

\newcommand{\Trh}{\ensuremath{T_{\rm rh}}}
\newcommand{\sg}{\ensuremath{\phi}}

\newcommand{\sgx}{\ensuremath{\phi_\star}}
\newcommand{\sgf}{\ensuremath{\phi_{\rm f}}}

\newcommand{\ld}{\ensuremath{\lambda}}
\newcommand{\ldu}{\ensuremath{\uplambda}}
\newcommand{\Ld}{\ensuremath{\Lambda}}
\newcommand{\kp}{\ensuremath{\kappa}}

\newcommand{\se}{\ensuremath{\widehat \phi}}
\newcommand{\sex}{\ensuremath{\widehat{\phi}_\star}}
\newcommand{\sef}{\ensuremath{\widehat{\phi}_{\rm f}}}
\newcommand{\geu}{\ensuremath{\widehat g}}
\newcommand{\eph}{\ensuremath{\widehat \epsilon}}
\newcommand{\ith}{\ensuremath{\widehat \eta}}

\newcommand{\mgr}{\ensuremath{m_{3/2}}}

\newcommand{\am}{\ensuremath{{\rm a}_{3/2}}}
\newcommand{\mg}{{\ensuremath{M_{1/2}}}}

\newcommand\mtta[4]{\mbox{
$\llgm\bem #1 &#2 \cr #3& #4\eem\rrgm$}}

\def\Kap{K\"{a}hler potential}

\def\sub{subplanckian}

\def\bcp{{\sc\small Bicep2}/{\it Keck Array}}
\newcommand{\plk}{{\it Planck}}

\newcommand{\diag}{\ensuremath{{\sf diag}}}
\renewcommand{\det}{\ensuremath{{\sf det}}}

\newcommand{\cm}{\ensuremath{c_{-}}}
\newcommand{\cp}{\ensuremath{c_{+}}}

\renewcommand{\refname}{{\bf\scshape References}}

\renewcommand{\thesubsection}{{\small\sf\Alph{subsection}}}

\renewenvironment{subequations}{%
\refstepcounter{equation}%
\setcounter{parentequation}{\value{equation}}%
  \setcounter{equation}{0}
  \ignorespaces
}{%
  \setcounter{equation}{\value{parentequation}}%
  \ignorespacesafterend
}

\begin{document}


\title{\boldmath\bf\scshape  Induced-Gravity GUT-Scale Higgs Inflation in Supergravity}

\author{\scshape Constantinos Pallis}
\affiliation{School of Electrical \& Computer Engineering, Faculty
of Engineering, Aristotle University of Thessaloniki, GR-541 24
Thessaloniki, GREECE \\  {\sl e-mail address: }{\ftn\tt
kpallis@gen.auth.gr}}
\author{\scshape  Qaisar Shafi}
\affiliation{Bartol Research Institute, Department of Physics and
Astronomy, \\ University of Delaware, Newark, DE 19716, USA\\
{\sl e-mail address: }{\ftn\tt shafi@bartol.udel.edu}}




\begin{abstract}

\noindent {\ftn \bf\scshape Abstract:} Models of induced-gravity
inflation are formulated within Supergravity employing as inflaton
the Higgs field which leads to a spontaneous breaking of a $\bl$
symmetry at $\Mgut=2\cdot10^{16}~\GeV$. We use a renormalizable
superpotential, fixed by a $U(1)$ R symmetry, and \Kap s which
exhibit a quadratic non-minimal coupling to gravity with or
without an independent kinetic mixing in the inflaton sector. In
both cases we find inflationary solutions of Starobinsky type
whereas in the latter case, others (more marginal) which resemble
those of linear inflation arise too. In all cases the inflaton
mass is predicted to be of the order of $10^{13}~\GeV$. Extending
the superpotential of the model with suitable terms, we show how
the MSSM $\mu$ parameter can be generated. Also, non-thermal
leptogenesis can be successfully realized, provided that the
gravitino is heavier than about $10~\TeV$.
\\ \\ {\scriptsize {\sf PACs numbers: 98.80.Cq, 04.50.Kd, 12.60.Jv, 04.65.+e}
\hfill {\sl\bfseries Published in} {\sl Eur. Phys. J. C} {\bf 78},
no. 6, 523 (2018).}

\end{abstract}\pagestyle{fancyplain}

\maketitle

\rhead[\fancyplain{}{ \bf \thepage}]{\fancyplain{}{\sl
Induced-Gravity GUT-Scale HI in SUGRA}} \lhead[\fancyplain{}{\sl
\leftmark}]{\fancyplain{}{\bf \thepage}} \cfoot{}

\section{Introduction}\label{intro}

The idea of \emph{induced gravity} ({\ftn\sf IG}), according to
which the (reduced) Planck mass $\mP$ is generated \cite{zee} via
the \emph{vacuum expectation value} ({\ftn\sf v.e.v}) that a
scalar field acquires at the end of a phase transition in the
early universe, has recently attracted  a fair amount of
attention. This is because it may follow an inflationary stage
driven by a Starobisky-type potential \cite{R2} in
\emph{Supergravity} ({\sf\ftn SUGRA}) \cite{R2r,nIG, su11, rena,
jones2} and in non-\emph{Supersymmetric} ({\sf\ftn SUSY})
\cite{ig,old,gian,jones,higgsflaton} settings, which turns out to
be nicely compatible with the observational data \cite{plin}. As a
bonus, the resulting effective theories do not suffer from any
problem with perturbative unitarity
\cite{cutoff,riotto,gian,R2r,nIG} in sharp contrast to some models
of non-minimal inflation \cite{nmi,atroest, oss, nmH} where the
inflaton after inflation assumes a v.e.v much smaller than $\mP$.

The simplest way to realize the idea of IG is to employ a
double-well potential, $\ld(\sg^2-v^2)^2$, for the inflaton $\sg$
\cite{zee, ig, old, gian, higgsflaton, R2r, rena, nIG} -- scale
invariant realizations of this idea are proposed in \cref{jones}.
If we adopt a non-minimal coupling to gravity
\cite{old,higgsflaton} of the type $\fr=\ca\sg^2$ and set
$v=\mP/\sqrt{\ca}$, then $\vev{\fr}=\mP^2$, i.e., $\fr$ reduces to
$\mP^2$ at the vacuum generating, thereby, Einstein gravity at low
energies. The implementation of inflation, on the other hand,
which requires the emergence of a sufficiently flat branch of the
potential at large field values constrains $\ca$ to sufficiently
large values and $\ld$ as a function of $\ca$. An even more
restrictive version of this scenario would be achieved if $\sg$ is
involved in a Higgs sector which triggers a \emph{Grand Unified
Theory} ({\sf\ftn GUT}) phase transition in the early Universe
\cite{old, jones2}. The scale of a such transition is usually
related to the (field dependent) mass of the lightest gauge boson
and can be linked to some unification condition in
\emph{supersymmetric} ({\sf\ftn SUSY}) -- most notably -- settings
\cite{nmH, jhep, nMHkin, var, univ}. As a consequence, $\ca$ can
be uniquely determined by the theoretical requirements, giving
rise to an economical, predictive and well-motivated set-up,
thereby called \emph{IG Higgs inflation} ({\sf\ftn IGHI}). To our
knowledge, the unification hypothesis has not been previously
employed in constraining IGHI.

Since gauge coupling unification is elegantly achieved within the
\emph{minimal supersymmetric standard model} ({\sf\ftn MSSM}), we
need to formulate IGHI in the context of SUGRA. Namely, we employ
a renormalizable superpotential, uniquely determined by a gauge
and a $U(1)$ R symmetry, which realizes the Higgs mechanism in a
SUSY framework. Actually, this is the same superpotential widely
used for the models of F-term hybrid inflation \cite{susyhybrid,
dvali, bl, gpp, okada}. Contrary to that case, though, where the
inflaton typically is a gauge singlet and a pair of gauge
non-singlets are stabilized at zero, here the inflaton is involved
in the Higgs sector of the theory whereas the gauge singlet
superfield is confined at the origin playing the role of a
\emph{stabilizer} -- for a related scenario see \cref{okadashafi}.
For this reason we call it \emph{Higgs inflation} ({\sf\ftn HI}).
As regards the \Kap s, $K$, we concentrate on semi-logarithmic
ones which  employ variable coefficients for the logarithmic part
and include only quadratic terms of the various fields, taking
advantage of the recently established \cite{su11} stabilization
mechanisms of the accompanying non-inflaton fields.

More specifically, we distinguish two different classes of $K$'s,
depending whether we introduce an independent kinetic mixing in
the inflaton sector or not. In the latter case the non-minimal
coupling to gravity reads $\fr\sim\ca\phi^{2}$ and imposing the IG
and unification conditions allows us to fully determine $\ca$. In
the former case, apart from the non-minimal coupling to gravity
expressed as $\fr=\cp\phi^{2}$, the models exhibit a kinetic
mixing of the form $\fk\simeq\cm\fr$, where the constants $\cm$
and $\cp$ can be interpreted as the coefficients of the principal
shift-symmetric term ($\cm$) and its violation ($\cp$) in the
$K$'s. Obviously these models are inspired by the kinetically
modified non-minimal HI studied in \cref{nMHkin, jhep, var, univ}.
The observables now depend on the ratio $r_\pm=\cp/\cm$ which can
be found precisely enforcing the IG and unification conditions. As
a consequence, for both classes of models more robust predictions
can be here achieved than those presented in the original papers
\cite{nmH, nMHkin, jhep, var, univ}, where $\mP$ is included in
$\fr$ from every beginning. Most notably, the level of the
predicted primordial gravitational waves is about an order of
magnitude lower than the present upped bound \cite{plin, gwsnew}
and may be detectable in the next generation of experiments
\cite{bcp3, prism, bird, core}.

We exemplify our proposal employing as ``GUT'' gauge symmetry
$\Gbl=G_{\rm SM}\times U(1)_{B-L}$, where ${G_{\rm SM}}=
SU(3)_{\rm C}\times SU(2)_{\rm L}\times U(1)_{Y}$ is the gauge
symmetry of the standard model, and $B$ and $L$ denote baryon and
lepton number respectively -- cf. \cref{bl,jhep, var, univ}. The
embedding of IGHI within this particle model gives us the
opportunity to connect inflation with low energy phenomenology. In
fact, the absence of the gauge anomalies enforces the presence of
three right-handed neutrinos $\sni$ which, in turn, generate the
tiny neutrino masses via the type I seesaw mechanism. Furthermore,
the out-of-equilibrium decay of the $\sni$'s provides us with an
explanation of the observed \emph{baryon asymmetry of the
universe} ({\ftn\sf BAU}) \cite{baryo} via \emph{non-thermal
leptogenesis} ({\sf\ftn nTL}) \cite{lept} consistently with the
gravitino ($\Gr$) constraint \cite{gravitino,brand,kohri,grNew}
and the data \cite{valle,lisi} on the neutrino oscillation
parameters. Also, taking advantage of the adopted $R$ symmetry,
the parameter $\mu$ appearing in the mixing term between the two
electroweak Higgs fields in the superpotential of MSSM is
explained as in \crefs{dvali, R2r, univ} via the v.e.v of the
stabilizer field, provided that the relevant coupling constant is
appropriately suppressed. The post-inflationary completion induces
more constraints testing further the viability of our models.

The remaining text is organized into three sections. We first
establish and analyze our inflationary scenarios in \Sref{hi}. We
then -- in \Sref{post} -- examine a possible post-inflationary
completion of our setting. Our conclusions are summarized in
\Sref{con}.  Throughout the text, the subscript of type $,z$
denotes derivation \emph{with respect to} ({\ftn\sf w.r.t}) the
field $z$, and charge conjugation is denoted by a star. Unless
otherwise stated, we use units where $\mP = 2.433\cdot
10^{18}~\GeV$ is taken to be unity.

\section{Inflationary Models}\label{hi}

In Sec.~\ref{hi1} we describe the generic formulation of IG models
within SUGRA, in \Sref{hiV}, we construct the inflationary
potential, and in \Sref{hian} we analyze the observational
consequences of the models.

\subsection{\small\sf\scshape Embedding Induced-Gravity HI in
SUGRA}\label{hi1}

The implementation of IGHI requires the determination of the
relevant super- and \Kap s, which are specified in \Sref{hiset}.
In \Sref{hivev} we present the form of the action in the two
relevant frames and in \Sref{hiig} we impose the IG constraint.

\subsubsection{\small\sf Set-up}\label{hiset}

As we already mentioned, we base the construction of our models on
the superpotential
\beq \Whi=\ld S\lf\bar\Phi\Phi-M^2/4\rg\label{Whi} \eeq
which is already introduced in the context of models of F-term
hybrid inflation \cite{susyhybrid}. Here $\bar{\Phi}$, $\Phi$
denote a pair of left-handed chiral superfields oppositely charged
under $U(1)_{B-L}$; $S$ is a $\Gbl$-singlet chiral superfield;
$\ld$ and $M$ are parameters which can be made positive by field
redefinitions. $\Whi$ is the most general renormalizable
superpotential consistent with a continuous R symmetry
\cite{susyhybrid} under which \beq S\  \to\
e^{i\alpha}\,S,~\bar\Phi\Phi\ \to\ \bar\Phi\Phi,~\Whi \to\
e^{i\alpha}\, \Whi\,.\label{Rsym} \eeq
Here and in the subsequent discussion the subscript HI is
frequently used instead of IGHI to simplify the notation.

As we verify below, $\Whi$ allows us to break the gauge symmetry
of the theory in a simple, elegant and restrictive way. The v.e.vs
of these fields, though, have to be related with the size of $\mP$
according to the IG requirement. To achieve this, together with
the establishment of an inflationary era, we have to combine
$\Whi$ with a judiciously selected \Kap, $K$. We present two
classes of such $K$'s, which respect the (gauge and global)
symmetries of $\Whi$ and incorporate only quadratic terms of the
various fields. We distinguish these classes taking into account
the origin of the kinetic mixing in the inflaton sector. Namely:
\subparagraph{\sl (a) $K$'s without independent kinetic mixing.}
Having in mind the general recipe \cite{linde1, nmH} for the
introduction of non-minimal couplings in SUGRA we include the
gauge invariant function \beq \hr=\bar\Phi\Phi\label{hr}\eeq in
the following $K$'s
%
\beqs\beq  K_{1\cal R}=-N\ln\lf\ca\lf\hr+
\hr^*\rg-\frac{|\Phi|^2+|\bar\Phi|^2}{N}+F_{1S}\rg,\label{K1r}\\
\eeq which is completely logarithmic, and \beq K_{2\cal
R}=-N\ln\lf\ca(\hr+\hr^*)-\frac{|\Phi|^2+|\bar\Phi|^2}{N}\rg+F_{2S}\,,\label{K2r}\eeq\eeqs
which is polylogarithmic. In both cases we take $N>0$. The crucial
difference of the $K$'s considered here, compared to those
employed in \cref{linde1, nmH}, is that unity does not accompany
the terms $\ca\lf\hr+\hr^*\rg$. As explained in \Sref{hiig}, the
identification of this quantity with unity at the vacuum of the
theory essentially encapsulates the IG hypothesis --
cf.~\cref{R2r,nIG}. The existence of the real function
$|\Phi|^2+|\bar\Phi|^2$ inside the argument of logarithm is vital
for this scenario, since otherwise the K\"ahler metric is
singular. These terms provide canonical kinetic terms for $K=\kar$
and $N=3$ in the Jordan frame or $\ca$-dependent kinetic mixing in
the remaining cases, as we show in the next Section.

\subparagraph{\sl (b) $K$'s with independent kinetic mixing.} In
this case we introduce a softly broken shift symmetry for the
Higgs fields -- cf. \cref{shiftHI, jhep} -- via the functions
$F_\pm=\left|\Phi\pm\bar\Phi^*\right|^2$. In particular, the
dominant shift symmetry adopted here is
\beq \Phi \to\ \Phi+c\>\>\>\mbox{and}\>\>\>\bar\Phi \to\
\bar\Phi+c^*\>\>\mbox{with}\>\>c\in\mathbb{C},\label{shift}\eeq
under which $\fm$ remains unaltered whereas $\fp$ expresses the
violation of this symmetry and is placed in the argument of a
logarithm with coefficient $(-N)$, whereas $\fm$ is set outside
it. Namely, we propose the following $K$'s
\beqs\bea
K_1&=&-N\ln\left(\cp\fp+F_{1S}(|S|^2)\right)+\cm\fm,\label{K1}\\
K_2&=&-N\ln\left(\cp\fp\right)+\cm\fm+F_{2S}(|S|^2),\label{K2}\\
K_3&=&-N\ln\left(\cp\fp\right)+F_{3S}(\fm, |S|^2), \label{K3}
\eea\eeqs
where $N>0$. As in the case of the $K$'s in \eqs{K1r}{K2r} unity
is not included in the argument of the logarithm. In the present
case, the identification of $\cp\fp$ with unity -- see \Sref{hiig}
-- at the vacuum of the theory incarnates the IG hypothesis --
cf.~\cref{R2r,nIG}. The degree of the violation of the symmetry in
\Eref{shift} is expressed by $\rs=\cp/\cm$, which is constrained
by the unification condition to values of the order $0.1$ -- see
\Sref{higut}. Since this value is quite natural we are not forced
here to invoke any argument regarding its naturalness -- cf.
\cref{univ}.

\subparagraph{} The models employing the $K$'s in \eqs{K1r}{K2r}
are more economical compared to the models based on the $K$'s in
Eqs.~(\ref{K1}) -- (\ref{K3}). Indeed, the latter include two
parameters ($\cp$ and $\cm$) from which one $(\cp)$ enters $\fr$
and the other ($\cm$) dominates independently the kinetic mixing
-- see below. However, these parameters are related to the shift
symmetry in \Eref{shift} which renders the relevant setting
theoretically more appealing. Indeed, this symmetry  has a string
theoretical origin as shown in \cref{lust}. In this framework,
mainly integer $N$'s are considered which can be reconciled with
the observational data -- see \Sref{hinum}. Namely, $N=3$ [$N=2$]
for $K=K_1$ [$K=K_2$ or $K_3$] yields completely acceptable
results. However, the deviation of the $N$'s from these integer
values is also acceptable \cite{roest, jhep, var, nIG, univ} and
assist us to cover the whole allowed domain of the observables.

Another possibility that could be inspected is what happens if we
place the term $\cm\fm$ inside the argument of the logarithm in
\eqs{K1}{K2} -- cf. \cref{jhep} -- considering the \Kap s
\beqs\bea
K_{01}&=&-N\ln\left(\cp\fp-\cm\fm/N+F_{1S}\right)\,,\label{K01}\\
K_{02}&=&-N\ln\left(\cp\fp-\cm\fm/N\right)+F_{2S}\,.\label{K02}\eea\eeqs
These $K$'s, though, reduce to $K_{1\cal R}$ and $K_{2\cal R}$
respectively if we set
\beq\cp=\frac{N\ca-1}{2N}~~\mbox{and}~~
\cm=\frac{N\ca+1}{2}.\label{cpm0}\eeq
For $\ca\gg1$ the arrangement above results in $\rs\simeq1/N$. On
the other hand, the same $\rs$ is found if we impose the
unification constraint. Therefore, the observational predictions
of the models based on the $K$'s above are expected to be very
similar to those obtained using \eqs{K1r}{K2r}.

The functions $F_{lS}$ with $l=1,2,3$ encountered in
Eqs.~(\ref{K1r}), (\ref{K2r}) and (\ref{K1}) -- (\ref{K3}) support
canonical normalization and safe stabilization of $S$ during and
after IGHI. Their possible forms are given in \cref{univ}. Just
for definiteness, we adopt here only their logarithmic form, i.e.,
\beqs\bea
F_{1S}&=&-\ln\left(1+|S|^2/N\right),\label{f1s}\\
F_{2S}&=&N_S\ln\left(1+|S|^2/N_S\right),\label{f2s}\\
F_{3S}&=&N_S\ln\left(1+|S|^2/N_S+\cm\fm/N_S\right), \label{f3s}
\eea\eeqs
with $0<N_S<6$. Recall \cite{linde1,su11} that the simplest term
$|S|^2$ leads to instabilities for $K=K_1$ and light excitations
for $K=K_2$ and $K_3$. The heaviness of these modes is required so
that the observed curvature perturbation is generated wholly by
our inflaton in accordance with the lack of any observational hint
\cite{plcp} for large non-Gaussianity in the cosmic microwave
background.

\subsubsection{\sf\small From Einstein to Jordan Frame}\label{hivev}

With the ingredients above we can extract the part of the
\emph{Einstein frame} ({\sf\ftn EF}) action within SUGRA related
to the complex scalars $z^\al=S,\phc,\phcb$ -- denoted by the same
superfield symbol. This has the form \cite{linde1}
\beqs\beq\label{Saction1}  {\sf S}=\int d^4x \sqrt{-\what{
\mathfrak{g}}}\lf-\frac{1}{2}\rce +K_{\al\bbet} \geu^{\mu\nu}D_\mu
z^\al D_\nu z^{*\bbet}-\Ve\rg\,, \eeq
where $\rce$ is the EF Ricci scalar curvature, $D_\mu$ is the
gauge covariant derivative, $K_{\al\bbet}=K_{,z^\al z^{*\bbet}}$,
and $K^{\al\bbet}K_{\bbet\gamma}=\delta^\al_{\gamma}$. Also, $\Ve$
is the EF SUGRA potential which can be found in terms of $\Whi$ in
\Eref{Whi} and the $K$'s in Eqs.~(\ref{K1}) -- (\ref{K3}) via the
formula
\beq \hspace*{-1.5mm}\Ve=e^{\Khi}\left(K^{\al\bbet}(D_\al W_{\rm
HI})D^*_\bbet W_{\rm HI}^*-3{\vert W_{\rm
HI}\vert^2}\right)+\frac{g^2}2 \mbox{$\sum_{\rm a}$} {\rm D}_{\rm
a}^2,\label{Vsugra} \eeq\eeqs
where $D_\al W_{\rm HI}=W_{{\rm HI},z^\al}+K_{,z^\al}W_{\rm HI}$,
${\rm D}_{\rm a}=z^\al\lf T_{\rm  a}\rg_\al^\bt K_\bt$ and the
summation is applied over the generators $T_{\rm a}$ of $\Gbl$. In
the \emph{right-hand side} ({\sf\ftn r.h.s}) of the equation above
we clearly recognize the contribution from the D terms
(proportional to $g^2$) and the remaining one which comes from the
F terms.

If we perform a conformal transformation, along the lines of
\cref{linde1, jhep}, defining the frame function as
\beq\label{omgdef}
-{\Omega/N}=\exp\lf-{K}/{N}\rg\>\Rightarrow\>K=-N
\ln\lf-\Omega/N\rg\,,\eeq
we can obtain the form of ${\sf S}$ in the \emph{Jordan Frame}
({\ftn\sf JF}) which is written as \cite{jhep}
\beqs\bea  \nonumber {\sf S}&=&\int d^4x
\sqrt{-\mathfrak{g}}\lf\frac{ \Omega}{2N}\rcc-
\frac{27}{N^3}\Omega{\cal A}_\mu{\cal A}^\mu-V+\right. \\ &&
\left.\lf\Omega_{\al{\bbet}}+\frac{3-N}{N}
\frac{\Omega_{\al}\Omega_{\bbet}}{\Omega}\rg D_\mu z^\al D^\mu
z^{*\bbet} \rg, \label{Sfinal}\eea where we use the shorthand
notation $\Omega_\al=\Omega_{,z^\al}$, and
$\Omega_\aal=\Omega_{,z^{*\aal}}$. We also set
$V=\Ve{\Omg^2}/{N^2}$ and
\beq {\cal A}_\mu =-iN \lf \Omega_\al D_\mu z^\al-\Omega_\aal
D_\mu z^{*\aal}\rg/6\Omega\,.\label{Acal}\eeq\eeqs
Computing the expression in the parenthesis of the second line in
\Eref{Sfinal} for $K=\kar$ and $\kbr$, we can easily verify that
the choice for $N=3$ ensures canonical kinetic terms -- in
accordance with the findings in \cref{linde1, nmH} -- whereas in
the remaining cases a $\ca$- (and not $\sg$-) dependent kinetic
mixing emerges. Indeed, in any case we have
$\Omega_{\al\bbet}=\delta_{\al\bbet}$ and for $N=3$ the second
term in the parenthesis vanishes. On the contrary, for $K=K_1,
K_2$ and $K_3$, the same expression is not only different than
$\delta_{\al\bbet}$ but also includes ($\sg$-dependent) entries
proportional to and dominated by $\cm\gg\cp$. For this reason, the
relevant models of IGHI may be more properly characterized as
kinetically modified. The non-renormalizability of this kinetic
mixing is under control since $\sg\ll1$ and the theory is
trustable up to $\mP$, as we show in \Sref{hiobs}.

Most importantly, though, the first term in the first line of the
r.h.s of \Eref{Sfinal} reveals that $-\Omega/N$ plays the role of
a non-minimal coupling to gravity. Comparing \Eref{omgdef} with
the $K$'s in Eqs.~(\ref{K1r}) -- (\ref{K3}) we can infer that
\beq
-\frac{\Omega}{N}=\begin{cases}2(N\ca+1)\hr/{N}&\mbox{for}~~K=\kar~~\mbox{and}~~\kbr,\\\cp\fp&
\mbox{for}~~K=K_1,K_2~~\mbox{and}~~K_3,
\end{cases}\label{minK1}\eeq
along the field configuration \beq
\Phi=\bar\Phi^*~~~\mbox{and}~~~S=0,\label{infrtr1}\eeq
which is a honest inflationary trajectory, as shown in
\Sref{hiV2}. The identification of the quantity in \Eref{minK1}
with $\mP^2$ at the vacuum, according to the IG conjecture, can be
accommodated as described in the next section.


\subsubsection{\sf\small Induced-Gravity
Requirement}\label{hiig}

The implementation of the IG scenario requires the generation of
$\mP$ at the vacuum of the theory, which thereby has to be
determined. To do this we have to compute $V$ in \Eref{Vsugra} for
small values of the various fields, expanding it in powers of
$1/\mP$. Namely, we obtain the following low-energy effective
potential
\beqs \beq \label{Vsusy} V_{\rm eff}= e^{\widetilde K}\widetilde
K^{\al\bbet} W_{\rm HI\al} W^*_{\rm HI\bbet}+\frac{g^2}2
\mbox{$\sum_{\rm a}$} {\rm D}_{\rm a}^2+\cdots,\eeq
where the ellipsis represents terms proportional to $\Whi$ or
$|\Whi|^2$ which obviously vanish along the path in \Eref{inftr1}
-- we assume here that the vacuum is contained in the inflarionary
trajectory. Also, $\widetilde K$ is the limit of the $K$'s in
Eqs.~(\ref{K1r}) -- (\ref{K3}) for $\mP\to\infty$. The absence of
unity in the arguments of the logarithms multiplied by $N$ in
these $K$'s prevents the drastic simplification of $\wtilde K$,
especially for $K=\kar$ and $K_1$ -- cf. \cref{univ}. As a
consequence, the expression of the resulting $V_{\rm eff}$ is
rather lengthy. For this reason we confine ourselves below to
$K=K_2$ or $K_3$ where $F_{lS}$ with $l=2, 3$ is placed outside
the first logarithm and so $\wtilde K$ can be significantly
simplified.  Namely, we get
\beq \label{Kquad}\widetilde K=-N\ln\cp\fp +\cm\fm
+|S|^2\,,\eeq\eeqs
from which we can then compute
\beqs\beq \lf \widetilde K_{\al\bbet}\rg=\diag\lf \widetilde
M_\pm,1\rg ~~\mbox{with}~~~\widetilde M_\pm=\mtta{\cm}{\widetilde
K_{\phc\phcb^*}}{\widetilde K_{\phc\phcb^*}}{\cm}.\label{Kab}\eeq
Here,
\beq \widetilde
K_{\phc\phcb^*}=\frac{N}{(\phc+\phcb^*)^2}~~~\mbox{and}~~~
\widetilde K_{\phcb\phc^*}=\frac{N}{(\phc^*+\phcb)^2}, \label{Ks2}
\eeq
since
\beq \widetilde
K_\phc=-{N}/({\phc+\phcb^*})+\cm(\phc^*-\phcb)\label{Ks11} \eeq
and \beq \widetilde
K_{\phcb}=-{N}/({\phc^*+\phcb})-\cm(\phc-\phcb^*)\,.\label{Ks12}
\eeq\eeqs
To compute $V_{\rm eff}$ we need to know
\beqs \beq\lf\wtilde K^{\al\bbet}\rg=\diag\lf \widetilde
M^{-1}_\pm,1\rg,\label{wkin}\eeq
where
\beq \widetilde M_\pm^{-1}=\frac{1}{\det\widetilde
M_\pm}\mtta{\cm}{-\widetilde K_{\phc\phcb^*}}{-\widetilde
K_{\phcb\phc^*}}{\cm},\label{Mpmi} \eeq
with
\beq  \det\widetilde M_\pm=\cm^2-N^2/\fp^2\,. \label{detMpm}
\eeq\eeqs
Upon substitution of \eqs{wkin}{Mpmi} into \Eref{Vsusy} we obtain
$$V_{\rm eff}\simeq\ld^2e^{\widetilde K_+}\left|\phcb\phc-\frac14{M^2}\right|^2+\frac{g^2}2\lf\phc
\widetilde K_{\phc}-\phcb \widetilde
K_{\phcb}\rg^2+$$\vspace*{-5mm}
\beq \frac{\ld^2e^{\widetilde K_+}|S|^2}{\det\widetilde
M_\pm}\lf\cm\lf|\phc|^2+|\phcb|^2\rg -\widetilde
K_{\phc\phcb^*}\phcb^*\phc-\widetilde
K_{\phcb\phc^*}\phcb\phc^*\rg, \label{VF}\eeq
where $\widetilde{K}_+=-N\ln\cp\fp$. We remark that the direction
in \Eref{inftr1} assures D-flatness since $\vev{\phc\widetilde
K_{\phc}}=\vev{\phcb\widetilde K_{\phcb}}$ and so the vacuum lies
along it with
\beq \vev{S}=0 \>\>\>\mbox{and}\>\>\>
|\vev{\Phi}|=|\vev{\bar\Phi}|=M/2\,.\label{vevs} \eeq
The same result holds also for $K=\kar,\kbr$ and $K_1$ as we can
verify after a more tedious computation. \Eref{vevs} means that
$\vev{\Phi}$ and $\vev{\bar\Phi}$ spontaneously break $U(1)_{B-L}$
down to $\mathbb{Z}^{B-L}_2$. Note that $U(1)_{B-L}$ is already
broken during IGHI and so no cosmic string are formed -- contrary
to what happens in the models of the standard F-term hybrid
inflation \cite{dvali, gpp}, which also employ $\Whi$ in
\Eref{Whi}.

Inserting \Eref{vevs} into \Eref{minK1} we deduce that the
conventional Einstein gravity can be recovered at the vacuum if
\beq
M=\begin{cases}\sqrt{{2N}/{(N\ca-1)}}&\mbox{for}~~K=\kar~~\mbox{and}~~\kbr,\\1/\sqrt{\cp}&
\mbox{for}~~K=K_1,K_2~~\mbox{and}~~K_3.
\end{cases} \label{ig}\eeq
For $\ca\simeq10^4$ or $\cp\sim(10^2-10^3)$ employed here, the
resulting values of $M$ are theoretically quite natural since they
lie close to unity. Indeed, since the form of $\Whi$ in \Eref{Whi}
is established around $\mP$ we expect that the scales entered by
hand in the theory have comparable size.

\subsection{\sf\scshape\small Inflationary Potential}\label{hiV}

Below we outline the derivation of the inflationary potential in
\Sref{hiV1} and check its stability by computing one-loop
corrections in \Sref{hiV2}. The last part of the analysis allows
us to determine the gauge-coupling unification condition (see
\Sref{higut}) which assists us to further constrain our models.

\subsubsection{\sf\small Tree-Level Result}\label{hiV1}

If we express $\Phi, \bar\Phi$ and $S$ according to the
parametrization
\beq\label{hpar} \Phi=\frac{\sg e^{i\th}}{\sqrt{2}}
\cos\thn,~~\bar\Phi=\frac{\sg e^{i\thb}}{\sqrt{2}}
\sin\thn~~~\mbox{and}~~~S=\frac{s +i\bar s}{\sqrt{2}}\,,\eeq
with $0\leq\thn\leq\pi/2$, the trough in \Eref{inftr1} can be
written as
\beq \label{inftr} \bar
s=s=\th=\thb=0\>\>\>\mbox{and}\>\>\>\thn={\pi/4}.\eeq
Along this the only surviving term in \Eref{Vsugra} is
\beqs\beq \label{Vhi0}\Vhi =e^{K}K^{SS^*}\, |W_{{\rm HI},S}|^2\,,
\eeq which, for the choices of $K$'s in Eqs.~(\ref{K1}) --
(\ref{K3}), reads \beq\Vhi=\frac{\ld^2\fw^2}{16{\rm
a}_W^2}\fr^{-N}\cdot\begin{cases}
%
\fr&\mbox{for}\>\> K=\kar, K_1,\\
1&\mbox{for}\>\> K=\kbr, K_2 ~~\mbox{and}~~
K_3,\end{cases}\label{Vhi1}\eeq
where $\fr^{-N}=e^{K}$ and we define the (inflationary) frame
function as
\beq \fr=-\left.{\Ohi\over
N}\right|_{\rm\Eref{inftr}}\label{minK0}\eeq which is translated
as
\beq\fr=\begin{cases}(N\ca-1)\sg^2/{2N}&\mbox{for}~~K=\kar~~\mbox{and}~~\kbr,
\\\cp\sg^2& \mbox{for}~~K=K_1,K_2~~\mbox{and}~~K_3.
\end{cases}\label{minK}\eeq
The last factor in \Eref{Vhi1} originates from the expression of
$K^{SS^*}$ for the various $K$'s. Also
\beq \fw=\begin{cases}
(N\ca-1)\sg^2-2N&\mbox{for}~~K=\kar~~\mbox{and}~~\kbr,\\
\cp\sg^2-1& \mbox{for}~~K=K_1,K_2~~\mbox{and}~~K_3,
\end{cases}\eeq
arises from the last factor in the r.h.s of \Eref{Vhi0} together
with ${\rm a}_W=(N\ca-1)$ for $K=\kar$ and $\kbr$ and ${\rm
a}_W=\cp$ for $K=K_1, K_2$ and $K_3$. If we set
\beq \label{ndef} N= \begin{cases} {2n+3} &\mbox{for}\>\> K=\kar,K_1, \\
2(n+1) &\mbox{for}\>\> K=\kbr,K_2 ~~\mbox{and}~~ K_3,
\end{cases}\eeq\eeqs
we arrive at a universal expression for $\Vhi$ which is \beq
\label{Vhi} \Vhi= \frac{\ld^2\fw^2}{16{\rm
a}_W^2\fr^{2(1+n)}}\,\cdot\eeq

\renewcommand{\arraystretch}{1.4}
\begin{table*}[!t]
\caption{\normalfont Mass-squared spectrum of the inflaton sector
for $K=\kar,\kbr,K_1, K_2$ and $K_3$  along the path in
\Eref{inftr}.}
\begin{ruledtabular}
\begin{tabular}{c|c|c|c|c||c|c|c}
{\sc Fields}&{\sc Eigen-} & \multicolumn{6}{c}{\sc Masses
Squared}\\\cline{3-8}
&{\sc States}&&$K=\kar$&$K=\kbr$& {$K=K_1$}&{$K=K_2$} & {$K=K_{3}$} \\
\colrule
4 Real &$\widehat\theta_{+}$&$\widehat
m_{\theta+}^2$&$6\Hhi^2(1-1/N)$ &{$6\Hhi^2$}
&\multicolumn{2}{c|}{$6\Hhi^2$}&{$6(1+1/N_S)\Hhi^2$}\\\cline{3-8}
Scalars&$\widehat \theta_\Phi$ &$\widehat m_{
\theta_\Phi}^2$&{$M^2_{BL}+6\Hhi^2\ca(N-3)$}&{$M^2_{BL}+6\Hhi^2\ca(N-2)$}&
\multicolumn{2}{c|}{$M^2_{BL}+6\Hhi^2$}&
{$M^2_{BL}+6(1+1/N_S)\Hhi^2$}\\\cline{3-8}
&$\widehat s, \widehat{\bar{s}}$ &$ \widehat
m_{s}^2$&$3\Hhi^2(N-6+\ca\phi^2/N)$&{$3\Hhi^2(4/N-4+N+2/N_S)$}&$6\fw\Hhi^2/N$
&\multicolumn{2}{c}{$6\Hhi^2/N_S$}\\\hline
1 Gauge&&&\multicolumn{2}{c||}{}&\multicolumn{3}{c}{}\\
Boson& $A_{BL}$ &$ M_{BL}^2$&\multicolumn{2}{c||}{$2Ng^2/\lf
N\ca-1\rg$}&\multicolumn{3}{c}{$g^2\lf\cm\sg^2-N\rg$}\\\colrule
$4$ Weyl & $\what \psi_\pm$ & $\what m^2_{\psi\pm}$
&\multicolumn{2}{c||}{$3\Hhi^2(\ca(N-2)\sg^2-2N)^2/N^2\ca^2\sg^4$}&
\multicolumn{3}{c}{$6\lf
N-\cp(N-2)\sg^2\rg^2\Hhi^2/\cm\fw^2\sg^2$}
\\\cline{3-8} Spinors &$\ldu_{BL}, \widehat\psi_{\Phi-}$&
$M_{BL}^2$&\multicolumn{2}{c||}{$2Ng^2/\lf
N\ca-1\rg$}&\multicolumn{3}{c}{$g^2\lf\cm\sg^2-N\rg$}
\end{tabular}\label{tab1}
\end{ruledtabular}
\end{table*}
%
\renewcommand{\arraystretch}{1.}

The value $n=0$ is special since we get $N=3$ for $K=\kar$ and
$K_1$ or $N=2$ for $K=\kbr$, $K_2$ or $K_3$. Therefore, $\Vhi$
develops an inflationary plateau as in the original case of
Starobinsky model within no-scale SUGRA \cite{R2r, su11} for large
$\ca$ or $\cp$. Contrary to that case, though, here we also have
$n$ and $\cm$, whose variation may have an important impact on the
observables -- cf. \cref{jhep, var}. In particular, for $n<0$,
$\Vhi$ remains an increasing function of $\sg$, whereas for $n>0$,
it develops a local maximum
\beq \Vhi(\sg_{\rm max})=\frac{\ld^2 n^{2 n}}{16{\rm a}^2(1 +
n)^{2 (1 + n)}}~~\mbox{at}~~ \sg_{\rm max}=\sqrt{\frac{1+n}{{\rm
a} n}}\,,\label{Vmax}\eeq
where ${\rm a}=\ca/2$ for $K=\kar$ and $\kbr$ whereas ${\rm
a}=\cp$ for $K=K_1, K_2$ and $K_3$. In a such case we are forced
to assume that hilltop \cite{lofti} IGHI occurs with $\sg$ rolling
from the region of the maximum down to smaller values. The
relevant tuning of the initial conditions can be quantified by
defining \cite{gpp} the quantity
\beq \Dex=\left(\sg_{\rm max} - \sgx\right)/\sg_{\rm
max}\,,\label{dms}\eeq
where $\sgx$ is the value of $\sg$ when the pivot scale
$\ks=0.05/{\rm Mpc}$ crosses outside the inflationary horizon. The
naturalness of the attainment of IGHI increases with $\Dex$, and
it is maximized when $\sg_{\rm max}\gg\sgx$ which results in
$\Dex\simeq1$.

To specify the EF canonically normalized inflaton, we note that,
for all choices of $K$ in Eqs.~(\ref{K1r}), (\ref{K2r}) and
(\ref{K1}) -- (\ref{K3}), $K_{\al\bbet}$ along the configuration
in \Eref{inftr} takes the form
\beq \lf K_{\al\bbet}\rg=\diag\lf M_\pm,K_{SS^*}\rg,\eeq where
$K_{SS^*}=1/\fr$ [$K_{SS^*}=1$] for $K=\kar, K_1$ [$K=\kbr, K_2$
and $K_3$]. For $K=\kar$ and $\kbr$ we find
\beq
M_\pm=\mtta{(1+N\ca)/2\fr}{N/\sg^2}{N/\sg^2}{(1+N\ca)/2\fr}\,.\label{Mpmr}
\eeq
and upon diagonalization we obtain the following eigenvalues
\beq
\label{kpmr}\kp_+=N\ca\fr^{-1}\>\>\>\mbox{and}\>\>\>\kp_-=\fr^{-1}.\eeq
Note that the existence of the real function
$|\Phi|^2+|\bar\Phi|^2$ inside the argument of logarithm is vital
for this scenario, since otherwise $M_\pm$ develops zero
eigenvalue and so it is singular, i.e., no $K^{\al\bbet}$ can be
defined. On the other hand, for $K=K_1, K_2$ and $K_3$ we obtain
\beq M_\pm=\mtta{\cm}{N/\sg^2}{N/\sg^2}{\cm},\label{Sni1} \eeq
with eigenvalues
\beq \label{kpm}\kp_\pm=\cm\pm N/\sg^2\,. \eeq
Given that the lowest $\sg$ value is given in \Eref{ig}, we can
impose, in this case, a robust restriction on the parameters to
assure the positivity of $\kp_-$ during and after IGHI. Namely,
\beq \label{kmpos}\kp_-\gtrsim0 ~~\Rightarrow
~~\rs\lesssim1/N\,,\eeq
whereas we are not obliged to impose any condition for $K=\kar$
and $\kbr$.

Inserting \eqs{hpar}{Sni1} in the second term of the r.h.s of
\Eref{Saction1} we can define the EF canonically normalized
fields, denoted by hat, as follows
\beqs\bea \label{Je} &&
\hspace*{-0.2cm}\frac{d\se}{d\sg}=J=\sqrt{\kp_+},\>\>\widehat{\theta}_+
={J\sg\theta_+\over\sqrt{2}},\>\> \widehat{\theta}_-
=\sqrt{\frac{\kp_-}{2}}\sg\theta_-\,,~~~~~~~~~~~~\\ && \label{Je1}
\hspace*{-0.2cm}\widehat \theta_\Phi =
\sg\sqrt{\kp_-}\lf\theta_\Phi-{\pi/4}\rg,~~(\what
s,\what{\bar{s}})=\sqrt{K_{SS^*}}(s,\bar s)\,,~~~\eea\eeqs
where $\th_{\pm}=\lf\bar\th\pm\th\rg/\sqrt{2}$. Note, in passing,
that the spinors $\psi_S$ and $\psi_{\Phi\pm}$ associated with the
superfields $S$ and $\Phi-\bar\Phi$ are similarly normalized,
i.e., $\what\psi_{S}=\sqrt{K_{SS^*}}\psi_{S}$ and
$\what\psi_{\Phi\pm}=\sqrt{\kp_\pm}\psi_{\Phi\pm}$ with
$\psi_{\Phi\pm}=(\psi_\Phi\pm\psi_{\bar\Phi})/\sqrt{2}$.

\subsubsection{\sf\small Stability and Loop-Corrections}\label{hiV2}

We can verify that the inflationary direction in \Eref{inftr} is
stable w.r.t the fluctuations of the non-inflaton fields. To this
end, we construct the mass-squared spectrum of the various scalars
defined in \eqs{Je}{Je1}. Taking the limit $\cm\gg\cp$ we find the
expressions of the masses squared $\what m^2_{\chi^\al}$ (with
$\chi^\al=\theta_+,\theta_\Phi$ and $S$) arranged in \Tref{tab1}.
For $\sg\simeq\sgx$ these fairly approach the quite lengthy, exact
expressions taken into account in our numerical computation. Given
that $\sg<0.1$ for $K=\kar$ and $\fw\gg1$ for $K=K_1$ we deduce
that $\what m^2_{s}>0$ for $N\simeq3$. Also for $K=\kbr$, $K_2$ or
$K_3$ and $0<N_S<6$, $\what m^2_{s}>0$ stays positive and heavy
enough, i.e. $\what m^2_{z^\al}\gg\Hhi^2=\Vhi/3$. In \Tref{tab1}
we also display the masses, $M_{BL}$, of the gauge boson $A_{BL}$
-- which signals the fact that $\Gbl$ is broken during IGHI -- and
the masses of the corresponding fermions. Note that the
unspecified eigestate $\what \psi_\pm$ is defined as \beq \what
\psi_\pm=\lf\what{\psi}_{\Phi+}\pm
\what{\psi}_{S}\rg/\sqrt{2}\,.\eeq
As a consequence, let us again emphasize that no cosmic string are
produced at the end of IGHI.

The derived mass spectrum can be employed in order to find the
one-loop radiative corrections, $\dV$, to $\Vhi$. Considering
SUGRA as an effective theory with cutoff scale equal to $\mP$, the
well-known Coleman-Weinberg formula \cite{cw} can be employed
taking into account only the masses which lie well below $\mP$,
i.e., all the masses arranged in \Tref{tab1} besides $M_{BL}$ and
$\what m_{\th_\Phi}$ -- note that these contributions are
cancelled out for $K=\kar$ and $N=3$ or $K=\kbr$ and $N=2$. The
resulting $\dV$ leaves intact our inflationary outputs, provided
that the renormalization-group mass scale $\Lambda$, is determined
by requiring $\dV(\sgx)=0$ or $\dV(\sgf)=0$. These conditions
yield $\Ld\simeq3.2\cdot10^{-5}-1.4\cdot10^{-4}$ and render our
results practically independent of $\Lambda$ since these can be
derived exclusively by using $\Vhi$ in \Eref{Vhi} with the various
quantities evaluated at $\Ld$ -- cf. \cref{jhep}. Note that their
renormalization-group running is expected to be negligible because
$\Ld$ is close to the inflationary scale
$\Vhi^{1/4}\simeq(3-7)\cdot10^{-3}$. Recall, here, that in the
case of F-term hybrid inflation \cite{susyhybrid, dvali, bl, gpp}
the SUSY potential is classically flat and the radiative
corrections contribute (together with the SUGRA corrections) in
the inclination of the inflationary path.

\subsubsection{\sf\small  SUSY Gauge Coupling
Unification}\label{higut}

The mass $M_{BL}$ listed in \Tref{tab1} of the gauge boson
$A_{BL}$ may, in principle, be a free parameter since the
$U(1)_{B-L}$ gauge symmetry does not disturb the unification of
the MSSM gauge coupling constants. To be more specific, though, we
prefer to determine $M_{BL}$ by requiring that it takes the value
$\Mgut$ dictated by this unification at the vacuum of the theory.
Namely, we impose
\beq \label{Mgut1}
\vev{M_{BL}}=\Mgut\simeq2/2.43\cdot10^{-2}=8.22\cdot10^{-3}\,.\eeq
This simple principle has important consequences for both classes
of models considered here. In particular:

\subparagraph{\sl (a) For $K=\kar$ or $\kbr$.} In this cases, the
condition above completely determines $\ca$ since it implies via
the findings of \Tref{tab1}
\beq \ca=\frac1N+\frac{2g^2}{\Mgut^2}\simeq1.451\cdot10^4\,,
\label{Mgutr}\eeq
leading to $M\simeq0.0117$ via \Eref{ig}. Here we take
$g\simeq0.7$ which is the value of the unified coupling constant
within MSSM. Although $\ca$ above is very large, there is no
problem with the validity of the effective theory, in accordance
with the results of earlier works \cite{R2r, gian, nIG} on IG
inflation with gauge singlet inflaton. Indeed, expanding about
$\vev{\phi}=M$ -- see \Eref{ig} -- the second term in the r.h.s of
\Eref{Saction1} for $\mu=\nu=0$ and $\Vhi$ in \Eref{Vhi} we obtain
\beqs\beq\label{Jexpr}  J^2 \dot\phi^2\simeq\lf 1
-\sqrt{\frac{2}{N}}\dphi+\frac{3}{2N}\dphi^2-\sqrt{\frac{2}{N^3}}\dphi^3
+\cdots\rg\dot{\what{\delta\sg}}^2,\eeq where $\dphi$ is the
canonically normalized inflaton at the vacuum -- see \Sref{lept1}
-- and \beq\Vhi\simeq\frac{\ld^2\dphi^2}{2N\ca^{2}}
\lf\hspace*{-0.1cm}1-\frac{2N-1}{\sqrt{2N}}\dphi+\frac{8N^2-4N+1}{8N}\what{\delta\sg}^2+\cdots
\hspace*{-0.1cm}\rg. \label{Vexpr}\eeq\eeqs
These expressions indicate that $\Qef=\mP$, since $\ca$ does not
appear in any of their numerators. Although these expansions are
valid only during reheating we consider $\Qef$ extracted this way
as the overall cut-off scale of the theory since reheating is
regarded \cite{riotto} as an unavoidable stage of IGHI.

\subparagraph{\sl (b) For $K=K_1, K_2$ or $K_3$.} In this cases,
the condition above allows us to fix $\rs$ since, substituting
\Eref{ig} in $M_{BL}$ shown in \Tref{tab1}, we obtain
\beq \label{Mgut}
g^2\lf\cm\vev{\sg}^2-N\rg=\Mgut^2\>\Rightarrow\>\rs=\frac{g^2}{Ng^2+\Mgut^2}\,.\eeq
Since $\Mgut>0$ the condition above satisfies the restriction in
\Eref{kmpos} yielding $\rs$ close to its upper bound because
$\Mgut\ll1$.

\subparagraph{} As a bottom line, under the assumption in
\Eref{Mgut1}, $\ca$ for $K=\kar$ and $\kbr$ or $\rs$ for $K=K_1,
K_2$ and $K_3$ cease to be free parameters, in sharp contrast to
the models of \cref{nmH, jhep, nMHkin, var, univ} where the same
assumption is employed to extract $M\ll1$ as a function of the
free parameters without any other theoretical constraint between
them. Therefore, the interplay of \eqs{ig}{Mgut} leads to the
reduction of the free parameters by one, thereby rendering the
present set-up more restrictive and predictive.

\subsection{\sf\scshape\small  Inflation Analysis}\label{hian}

In \Srefs{hiobs} and \ref{hinum} below we inspect analytically and
numerically respectively, if the potential in \Eref{Vhi} endowed
with the condition of \eqs{ig}{Mgut} may be consistent with a
number of observational constraints introduced in
Sec.~\ref{hiobs0}.

\subsubsection{\sf\small  General Framework}\label{hiobs0}

Given that the analysis of inflation in both EF and JF yields
equivalent results \cite{old}, we carry it out exclusively in the
EF. In particular, the period of slow-roll IGHI is determined in
the EF by the condition
\beqs\beq{\ftn\sf
max}\{\eph(\phi),|\ith(\phi)|\}\leq1,\label{srcon}\eeq where the
slow-roll parameters \cite{review},
\beq\label{sr}\widehat\epsilon= \left({\Ve_{\rm
HI,\se}/\sqrt{2}\Ve_{\rm HI}}\right)^2
\>\>\>\mbox{and}\>\>\>\>\>\widehat\eta={\Ve_{\rm
HI,\se\se}/\Ve_{\rm HI}}. \eeq\eeqs The number of e-foldings $\Ns$
that the scale $\ks=0.05/{\rm Mpc}$ experiences during  IGHI and
the amplitude $\As$ of the power spectrum of the curvature
perturbations generated by $\sg$ can be computed using the
standard formulae  \cite{review}
\begin{equation}
\label{Nhi}  \Ns=\int_{\sef}^{\sex} d\se\frac{\Vhi}{\Ve_{\rm
HI,\se}}~~~\mbox{and}~~~\As^{1/2}= \frac{1}{2\sqrt{3}\, \pi} \;
\frac{\Vhi^{3/2}(\sex)}{|\Ve_{\rm HI,\se}(\sex)|},\eeq
where $\sgx~[\sex]$ is the value of $\sg~[\se]$ when $\ks$ crosses
the inflationary horizon. These observables are to be confronted
with the requirements \cite{plcp}
\beqs \bea \label{Ntot}
\Ns&\simeq&61.5+\ln{\Vhi(\sgx)^{1/2}\over\Vhi(\sgf)^{1/4}}+\frac12\fr(\sgx);\\
\label{prob} \As^{1/2}&\simeq&4.627\cdot10^{-5}\,.\eea\eeqs
Note that in \Eref{Ntot} we consider an equation-of-state
parameter $w_{\rm int}=1/3$ corresponding to quartic potential
which is expected to approximate $\Vhi$ rather well for $\sg\ll1$
-- see \cref{jhep}. We obtain $\Ns\simeq(57.5-60)$.

Then, we compute the remaining inflationary observables, i.e., the
(scalar) spectral index $n_{\rm s}$, its running $a_{\rm s}$, and
the scalar-to-tensor ratio $r$ which are found from the relations
\cite{review}
\beqs\bea \label{ns} && \ns=\: 1-6\widehat\epsilon_\star\ +\
2\widehat\eta_\star,~~r=16\widehat\epsilon_\star, \\ \label{as} &&
\as =\:2\left(4\widehat\eta_\star^2-(n_{\rm
s}-1)^2\right)/3-2\widehat\xi_\star, \eea\eeqs
where the variables with subscript $\star$ are evaluated at
$\sg=\sgx$ and $\widehat\xi={\Ve_{\rm HI,\widehat\phi} \Ve_{\rm
HI,\widehat\phi\widehat\phi\widehat\phi}/\Ve^2_{\rm HI}}$.

The resulting values of $\ns$ and $r$ must be in agreement with
the fitting of the data \cite{plin,gwsnew} with $\Lambda$CDM$+r$
model. We take into account the data from \plk\ and \emph{Baryon
Acoustic Oscillations} ({\sf\ftn BAO}) and the {\sf\ftn BK14} data
taken by the \bcp\ CMB polarization experiments up to and
including the 2014 observing season. The results are
\begin{equation}  \label{nswmap}
\mbox{\ftn\sf
(a)}\>\>\ns=0.968\pm0.009\>\>\>~\mbox{and}\>\>\>~\mbox{\ftn\sf
(b)}\>\>r\leq0.07,
\end{equation}
at 95$\%$ \emph{confidence level} ({\sf\ftn c.l.}) with
$|\as|\ll0.01$.

\subsubsection{\sf\small  Analytic Results}\label{hiobs}

A crucial difference of the present analysis w.r.t that for the
models in \cref{nmH, jhep, nMHkin, var, univ} is that $M$, given
by \Eref{ig}, is not negligible during the inflationary period and
enters the relevant formulas via the function $\fw$ defined below
\Eref{Vhi1}. We find it convenient to expose separately our
results for the two basic classes of models introduced in
\Sref{hiset}. Namely:
\subparagraph{\sl (a) For $K=\kar$ and $\kbr$.}%
The slow-roll parameters can be derived employing $J$ in
\Eref{Vhi}, without explicitly expressing $\Vhi$ in terms of
$\se$. Our results are
\beq\label{srr}\eph=4\frac{\tfw^2(n\tfw-2)^2}{N\ca^4\sg^8}\>\>\mbox{and}\>\>
\ith=8\frac{2-\tfw-4 n\tfw+n^2\tfw^2}{N \tfw^{2}},\eeq
where $\tfw=\ca\sg^2-2$. The condition \Eref{srcon} is violated
for $\sg=\sgf$, which is found to be
\beq \sg_{\rm f}\simeq \mbox{\sf\small max}
\lf\frac2{\sqrt{\ca}}\sqrt{\frac{1+n}{2n+\sqrt{N}}},
2\sqrt{\frac2\ca}\sqrt{\frac{1-4n}{8n^2+N}}\rg\,.\label{sgfr}\eeq
Then, $\Ns$ can be also computed from \Eref{Nhi} as follows
\begin{equation}
\label{Nhianr}  \Ns\simeq\begin{cases}
N\ca\sgx^2/8&\mbox{for}~~n=0\,,
\\ N\ln\lf \frac{2(1+n)}{2-n\tilde f_{W\ast}}\rg/4 n(1+n) &\mbox{for}~~n\neq0\,,
\end{cases}
\end{equation}
where $\tilde f_{W\ast}=\tfw(\sex)$. Solving the above equations
w.r.t $\sgx$ we obtain a unified expression
\begin{equation} \label{sgxbr}\sgx\simeq\sqrt{\frac{2\frs}{\ca}}\>\>\>
\mbox{with}\>\>\>\frs=\frac{1+n}{n}\lf1-e^{-4n(1+n)\Ns/N}\rg
\end{equation}
reducing to $4\Ns/N$ in the limit $n\to0$. For $\ca$ in
\Eref{Mgutr} we can verify that $\sgx\sim0.1$ and so the model is
(automatically) well stabilized against corrections from higher
order terms of the form $(\phc\phcb)^p$ with $p>1$ in $\Whi$ --
see \Eref{Whi}. Thanks to \Eref{Mgutr}, we can derive uniquely
$\ld$ from the expression
\beq \ld=8\sqrt{6\As}\pi
\ca\frs^{n+1}\frac{n(1-\frs)+1}{\sqrt{N}(\frs-1)^{2}}\,,
\label{lanr}\eeq
applying the second equation in \Eref{Nhi}. Upon substitution of
$\frs$ into \Eref{ns} we obtain the predictions of the model which
are
\beqs\bea\label{nsr}&& \hspace*{-8mm}\ns\simeq1 - \frac{8n^2}{N}+
\frac{16}{N}\frac{n}{\frs-1}-\frac8N\frac{\frs+1}{(\frs-1)^2},\\
&&\hspace*{-8mm}r\simeq\frac{64}{N}\lf\frac{1+n(1-\frs)}{\frs-1}\rg^2\,.\label{rs1r}
\eea\eeqs
Since only $|n|\ll1$ are allowed, as we see below, the results
above, together with $\as$, can be further simplified as follows
\beqs\bea \label{ns2r} \ns&\simeq&
1-\frac{2}{\Ns}-\frac{4n}{N}-\frac{8n^2\Ns}{3N^2}\,,\\
\label{rs2r} r&\simeq&
\frac{4N}{\Ns^2}-\frac{16n^3}{N}+\frac{80n^2}{3N}-\frac{64n^3\Ns}{3N^2}\,,\\
\label{as2r}
\as&\simeq&-\frac{2}{\Ns^{2}}+\frac{3n}{\Ns^2}+\frac{8n^2}{3N^2}-\frac{7N}{2\Ns^3}\,,\eea\eeqs
where, for $n=0$, the well-known predictions of the Starobinsky
model are recovered, i.e., $\ns\simeq0.966$ and $r=0.0032$
[$r=0.0022$] for $K=\kar$ [$K=\kbr$]. On the other hand,
contributions proportional to $\Ns$ can be tamed for sufficiently
low $n$ as we can verify numerically.

\subparagraph{\sl (b) For $K=K_1, K_2$ and $K_3$.} Working along
the lines of the previous paragraph we estimate the slow-roll
parameters as follows
\beqs\bea\label{sr1}&& \eph=\frac{8(1+n-n\cp\sg^2)^2}{\cm\sg^2 \fw^{2}};\\
&&
\frac{\ith}{4}=\frac{5+9n-(3+10n)\cp\sg^2+4n^2\fw^2+n\cp^2\sg^2}{\cm\sg^2
\fw^{2}} \label{sr2}\cdot~~~~~~~~\eea\eeqs
Given that $\sg\ll1$, \Eref{srcon} is saturated at the maximal
$\sg$ value, $\sgf$, from the following two values
\beq \sg_{1\rm
f}\simeq\sqrt{\frac2\cm}\frac1{\rs^{1/3}}\>\>\>\mbox{and}\>\>\>
\sg_{2\rm
f}\simeq\sqrt{\frac2\cm}\lf\frac3\rs\rg^{1/4},\label{sgf}\eeq
where $\sg_{1\rm f}$ and  $\sg_{2\rm f}$ are such that
$\eph\lf\sg_{1\rm f}\rg\simeq1$ and $\ith\lf\sg_{2\rm
f}\rg\simeq1$. The $n$ dependence is not so crucial for this
estimation. Since $\sgx\gg\sgf$, from \Eref{Nhi} we find
\begin{equation}
\label{Nhian}  \Ns\simeq\begin{cases}
\cm\sgx^2\lf\cm\rs\sgx^2/2-1\rg/8&\hspace*{-2mm}\mbox{for}~~n=0,
\\ -\lf n \cp \sgx^2 +\ln\lf1- \frac{n \cp\sgx^2}{1+n}\rg\rg/8 n^2
\rs&\hspace*{-2mm}\mbox{for}~~n\neq0,
\end{cases}
\end{equation}
where $\sex$ is the value of $\se$ when $\ks$ crosses the
inflationary horizon. As regards the consistency of the relation
above for $n>0$, we note that we get $n \cp\sgx^2<1+n$ in all
relevant cases and so, $\ln(1 - n \cp\sgx^2/(1+n))<0$ assures the
positivity of $\Ns$. Solving the equations above w.r.t $\sgx$, we
can express $\sgx$ in terms of $\Ns$ as follows
\beq \label{sgxb}\hspace*{-0mm}\sgx\simeq
\frac{\frs^{\frac12}}{\cp^{\frac12}}~~\mbox{with}~~\frs=\begin{cases}
1+\lf1+16\rs\Ns\rg^{\frac12}&\hspace*{-2mm}\mbox{for}~~n=0,\\
\lf1+n+W_k(y)\rg/n&\hspace*{-2mm}\mbox{for}~~n\neq0,
\end{cases}
\end{equation}
where we make use of \Eref{minK}. Also, $W_k$ is the Lambert $W$
(or product logarithmic) function \cite{wolfram} with \beq
y=-(1+n)\exp\lf-1-n(1+8n\Ns\rs)\rg.\eeq We take $k=0$ for $n>0$
and $k=-1$ for $n<0$.

Contrary to what happens for $K=\kar$ and $\kbr$, $\cm$ is not
uniquely determined here. Therefore, for any $n$ we are obliged to
impose a lower bound on it, above which $\sgx\leq1$. Indeed, from
\Eref{sgxb} we have
\begin{equation}
\label{cmmin}\sgx\leq1~~\Rightarrow~~\cm\geq{\frs}/{\rs},
\end{equation}
and so our proposal can be stabilized against corrections from
higher order terms. Despite the fact that $\cm$ may take
relatively large values, the corresponding effective theories are
valid \cite{cutoff, riotto} up to $\mP=1$ for $\rs$ given by
\Eref{Mgut}. To further clarify this point we have to identify the
ultraviolet cut-off scale $\Qef$ of the theory by analyzing the
small-field behavior of our models.  More specifically, adapting
the expansions in \eqs{Jexpr}{Vexpr} in our present case, we end
up with the expressions
\beqs\beq\label{Jexp}  J^2 \dot\phi^2\simeq\lf1-2\rsb^3\dphi
+3N\rsb^4\what{\delta\sg}^2-4N\rsb^5\what{\delta\sg}^3+\cdots\rg\dot{\what{\delta\sg}}^2,\eeq
where we set $\rsb=\sqrt{\rs/(1+N\rs)}$, and
\bea\nonumber\Vhi&\simeq&\frac{\ld^2\rsb^2\dphi^2}{4\cp^{2}}\lf1-(3+4n)\rsb\dphi\right.\\
&&\left.+\lf\frac{25}{4}+14n+8n^2\rg\rsb^2\what{\delta\sg}^2+\cdots\rg.
\label{Vexp}\eea\eeqs
From the expressions above we conclude that $\Qef=\mP$ since
$\rs\leq1$ (and so $\rsb\leq1$) due to \Eref{Mgut}.

From the second equation in \Eref{Nhi} we can also conclude that
$\ld$ is proportional to $\cm$ for fixed $n$. Indeed, plugging
\Eref{sgxb} into this equation and solving w.r.t $\ld$, we find
\beq \ld=32\sqrt{3\As}\pi
\cm\rs^{3/2}\frs^{n+1/2}\frac{n(1-\frs)+1}{(\frs-1)^{2}}\,.\label{lan}\eeq
Numerically, -- see below -- we find that $\ld/\cm$ develops a
maximum at $n\simeq-0.15$ which signals a transition to a branch
of inflationary solutions which deviate from those obtained within
the Starobinsky-like inflation.

Inserting $\frs$ from \Eref{sgxb} into \eqs{ns}{as} we obtain
\beqs\bea\label{ns1}&&\hspace*{-7mm}  \ns\simeq1 - \frac8\frs
\lf\frac{3\frs+1}{(\frs-1)^2}-n\frac{\frs+3}{\frs-1}+2n^2\rg,\\
&&\hspace*{-7mm}r\simeq128\frac\rs\frs\lf\frac{1-n(\frs-1)}{\frs-1}\rg^2,\label{rs1}\\
&&\hspace*{-7mm}\as\simeq \frac{64 \rs^2}{3 (\frs-1)^4 \frs^2}
\Big(3 - 9 \frs (2 \frs+ 1 )\nonumber  \\&& \hspace*{-7mm} +\ 3
(\frs-1)(\frs (7 \frs+ 9 )-4) n \nonumber \\&& \hspace*{-7mm}  +\
2 (\frs-1)^2 (\frs(\frs-42) + 121 ) n^2\Big)\,.
\label{as1}\eea\eeqs
where we can recognize the similarities with the formulas given in
\eqs{nsr}{rs1r}. For $|n|<0.1$ these formulas may be expanded
successively in series of $n$ and $1/\Ns$ with results
\beqs\beq \label{ns2} \ns\simeq
1-\frac{16}{3}n^2\rs-2n\frac{\rs^{1/2}}{\Ns^{1/2}}-\frac{3-2n}{2\Ns}-\frac{3+5n}{24(\Ns^3\rs)^{1/2}}\,,\eeq
\vspace*{-5mm}\beq \label{rs2} r\simeq
-\frac{8n}{\Ns}-\frac{1}{2\Ns^2\rs}+\frac{2(3+2n)}{3(\Ns^3\rs)^{1/2}}
+\frac{32n^2\rs^{1/2}}{3\Ns^{1/2}}\,, \eeq \vspace*{-5mm}\beq
\label{as2}
\as\simeq-\frac{n\rs^{1/2}}{\Ns^{3/2}}-\frac{3-2n}{2\Ns^2}\,.\eeq\eeqs
From the expressions above, we can infer that there is a clear $n$
(and $\rs$) dependence of the observables which deviate somewhat
from those obtained in the pure Starobinsky-type inflation (or IG
inflation) \cite{R2r,nIG,su11}. Note that the formulae, although
similar, are not identical with those found in \cref{univ}.

\subsubsection{\sf\small  Numerical Results}\label{hinum}

The approximate analytic expressions above can be verified by the
numerical analysis. Namely, we apply the accurate expressions in
\Eref{Nhi} and confront them with the requirements in
\Erefs{Ntot}{prob} adjusting $\ca$ and $\ld$ for $K=\kar$ and
$\kbr$ or $\cm$ and $\ld$ for  with any selected $n$. Then, we
compute the model predictions via \eqs{ns}{as}. Our results are
mainly displayed in \Frefs{fig1r} and \ref{fig1}, where we show a
comparison of the models' predictions against the observational
data \cite{plin,gwsnew} in the $\ns-\rw$ plane, where
$\rw=16\eph(\se_{0.002})$ with $\se_{0.002}$ being the value of
$\se$ when the scale $k=0.002/{\rm Mpc}$, which undergoes $\what
N_{0.002}=\Ns+3.22$ e-foldings during IGHI, crosses the horizon of
IGHI. Let us discuss separately the results for the two classes of
models. In particular:

\begin{figure}[!t]
\centering
\includegraphics[width=60mm,angle=-90]{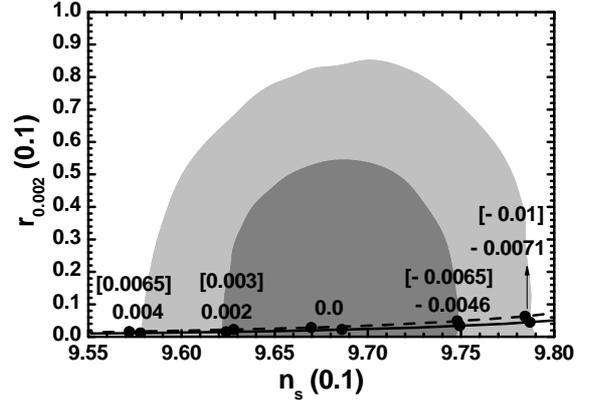}
\caption{\sl Allowed curves in the $\ns-\rw$ plane for $K=\kar$
(dashed line) and $K=\kbr$ (solid line) -- the $n$ values in
[outside] squared brackets correspond to $K=\kar$ [$K=\kbr$]. The
marginalized joint $68\%$ [$95\%$] regions from \plk, BAO and BK14
data are depicted by the dark [light] shaded
contours.}\label{fig1r}
\end{figure}

\subparagraph{\sl (a) For $K=\kar$ and $\kbr$.}  We depict in
\Fref{fig1r} by a dashed [solid] line the model predictions for
$K=\kar$ [$K=\kbr$] against the observational data. We see that
the whole observationally favored range at low $r$'s is covered
varying $n$ which remains, though, rather close to zero. In fact
$n$ is tuned closer to zero and $r$ is slightly lower compared to
those obtained for $K=K_1,K_2$ and $K_3$ -- see below. More
explicitly, we find the allowed ranges
\beqs\beq\label{res1r}
0.9\gtrsim{n}/{0.01}\gtrsim-1\>\>\>\mbox{and}\>\>\> 1.5\lesssim
{r}/{10^{-3}}\lesssim6.6\eeq for $K=\kar$, whereas for $K=\kbr$ we
have \beq \label{res2r} 5.1\gtrsim
{n/0.001}\gtrsim-9\>\>\>\mbox{and}\>\>\> 1.1\lesssim
{r/10^{-3}}\lesssim5.9\,. \eeq\eeqs
As $n$ varies in its allowed ranges presented below, we obtain
\beq 2.3\lesssim\ld/{0.1}\lesssim4
\>\>\mbox{or}\>\>1.9\lesssim\ld/{0.1}\lesssim3.5, \eeq
for $K=\kar$ or $K=\kbr$ respectively. If we take $n=0$, we find
the central values of $\ld$ in the ranges above which are $0.29$
and $0.24$ correspondingly.

\subparagraph{\sl (b) For $K=K_1, K_2$ and $K_3$.} In this case,
let us clarify that the (theoretically) free parameters of our
models are $n$ and $\ld/\cm$ and not $n$, $\cm$, and $\ld$ as
naively expected -- recall that $M$ and $\rs$ are found from
\eqs{ig}{Mgut}. Indeed, if we perform the rescalings
\beq\label{resc}
\Phi\to\Phi/\sqrt{\cm},~~\bar\Phi\to\bar\Phi/\sqrt{\cm}~~~\mbox{and}~~~
S\to S,\eeq
$\Whi$ in \Eref{Whi} depends on $\ld/\cm$  and $\rs^{-1}$, while
the $K$'s in \Eref{K1} -- (\ref{K3}) depend on $n$ and $\rs$. As a
consequence, $\Vhi$ depends exclusively on $\ld/\cm$ and $n$.
Since the $\ld/\cm$ variation is rather trivial -- see \Eref{lan}
-- we focus below on the variation of $n$.

\begin{figure}[!t]
\centering
\includegraphics[width=60mm,angle=-90]{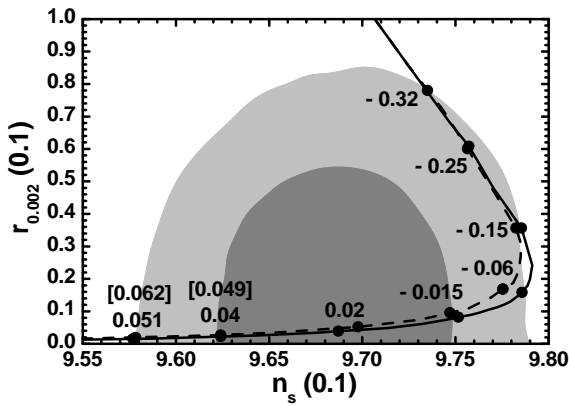}
\caption{\sl The same as \Fref{fig1r} but for $K=K_1$ (dashed
line) and $K=K_2$ or $K_3$ (solid line) with the $n$ values
indicated on the curves (the $n$ values in squared brackets
correspond to $K=K_1$). }\label{fig1}
\end{figure}

In \Fref{fig1} we depict the theoretically allowed values with
solid and dashed lines for $K=K_2$ or $K_3$ and $K=K_1$
respectively. The variation of $n$ is shown along each line. In
squared brackets we display the $n$ values for $K=K_1$ when these
differ appreciably from those for $K=K_2$ or $K_3$. We remark that
for $n>0$ there is a discrepancy of about $20\%$ changing $K$ from
$K_1$ to $K_2$ or $K_3$, which decreases as $n$ decreases below
zero. This effect originates from the difference in $J$ -- see
\eqs{kpm}{Je} -- which becomes smaller and smaller as $n$
decreases or $\cm$ increases. We observe that $n>0$ values
dominate the part of the curves with lower $r$ values and
$\ns\leq0.973$, whereas the $n<0$ values generate the part of the
curves with $\ns$ close to its upper bound in \Eref{nswmap} and
appreciably larger $r$ values. Roughly speaking, the displayed
curves can be produced interconnecting the limiting points of the
various curves in Fig.~2-{\ftn\sf (a)} of \cref{univ}, although
the curves for $0<n<0.1$ and $n<-0.1$ are not depicted there. This
is because the $\rs$'s resulting from \Eref{Mgut} are close to
their upper limits induced by \Eref{kmpos}.

\begin{table}[!t]
\caption{\normalfont Parameters and observables for the points
shown in \Fref{fig1} with $K=K_2$ and $K_3$.}
\begin{ruledtabular}
\begin{tabular}{c||c|c||c|c|c}
$n/0.1$&$\rs/0.1$&$\ld/10^{-5}\cm$&$\ns/0.1$&$-\as/10^{-4}$&$r/0.01$\\\colrule
$0.51$&$4.76$&$1.7$&$9.58$&$5.2$&$0.17$\\
$0.4$&$4.81$&$2$&$9.62$&$5.2$&$0.25$\\\colrule
$0.2$&$4.9$&$2.55$&$9.69$&$5$&$0.43$\\
$-0.15$&$5.07$&$3.4$&$9.75$&$4.5$&$0.88$\\
$-0.6$&$5.32$&$4.1$&$9.78$&$3.5$&$1.7$\\\colrule
$-1.5$&$5.88$&$4.5$&$9.78$&$3.98$&$3.7$\\
$-2.5$&$6.66$&$3.9$&$9.76$&$4$&$6.3$\\
$-3.2$&$7.35$&$3.3$&$9.73$&$4.4$&$8.3$\\
\end{tabular}
\end{ruledtabular}
\label{tab5}
\end{table}

Comparing these theoretical outputs with data depicted by the dark
[light] shaded contours at $68\%$ c.l. [$95\%$ c.l.] we find the
allowed ranges. Especially, for $K=K_1$ we obtain
\beq\label{res1} 0.62\gtrsim{n/0.1}\gtrsim-3.2,\>\>\> 3.2\lesssim
{\rs/0.1}\lesssim4.16.\eeq On the other hand, for $K=K_2$ or
$K_3$, we find one branch localized in the ranges
\beq\label{res21} 0.51\gtrsim{n/0.1}\gtrsim-0.6,\>\>\>
4.76\lesssim {\rs/0.1}\lesssim5.32,\eeq and another one
\beq\label{res22} -1.5\gtrsim {n/0.1}\gtrsim-3.2,\>\>\>
5.88\lesssim {\rs/0.1}\lesssim7.35.\eeq The findings for $K=K_2$
or $K_3$, can also be read-off from \Tref{tab5} where we list the
values of the input parameter ($n$) depicted in \Fref{fig1}, the
corresponding output parameters ($\rs$ and $\ld/\cm$) and the
inflationary observables. We observe that $\ns$ and $r$ are well
confined in the allowed regions of \Eref{nswmap}, while $\as$
varies in the range $-(3.98-5.2)\cdot10^{-4}$ and so, our models
are consistent with the fitting of data with the $\Lambda$CDM+$r$
model \cite{plin}. Comparing these numerical values with those
obtained by the analytic expressions in Eqs.~(\ref{ns1}) --
(\ref{as1}) we obtain complete agreement for any $n$. On the other
hand, the approximate formulas in Eqs.~(\ref{ns2}) -- (\ref{as2})
are valid only for $|\ns|<0.1$, i.e., the Starobinsky-like region.
Hilltop IGHI is attained for $n>0$ and there we find
$\Dex\gtrsim0.155$, where $\Dex$ increases as $n$ drops. The
required tuning is not severe, mainly for $n<0.04$ since
$\Dex\gtrsim20\%$. Since our models predict $r\gtrsim0.0017$, they
are testable by the forthcoming experiments, like {\sc Bicep3}
\cite{bcp3}, PRISM \cite{prism}, LiteBIRD \cite{bird} and CORE
\cite{core}, which are expected to measure $r$ with an accuracy of
$10^{-3}$. We do not present in \Tref{tab5} $\sgx$ values since,
as inferred by \Eref{lan}, every $\sgx$ satisfying \Eref{Ntot}
leads to the same ratio $\ld/\cm$. For the reasons mentioned below
\Eref{cmmin}, we prefer $\sgx\leq1$. To achieve this, we need
$\cm\gtrsim(30-140)$ for $K=K_1$ and $\cm\gtrsim(40-160)$ for
$K=K_2$ or $K_3$, where the variation of $\cm$ is given as $n$
decreases.

\begin{figure}[!t]
\centering
\includegraphics[width=60mm,angle=-90]{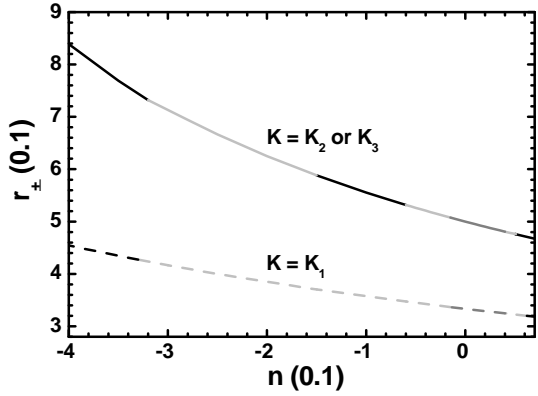}
\caption{\sl Values of $\rs$ allowed by \Eref{ig}  as a function
of $n$ for $K=K_1$ (dashed lines) and $K=K_2$ or $K_3$ (solid
lines). The values which are also preferred by the observational
data at $68\%$ c.l. [$95\%$ c.l.] are included in the dark [light]
gray segments.}\label{nrs}
\end{figure}


For $K=K_1$ we expect similar values for $\ld/\cm$ and the
inflationary observables. However, $\rs$ will differ appreciably
due to the different relation between $n$ and $N$ -- see
\Eref{ndef}. To highlight it further, we present in \Fref{nrs} the
$\rs$ values, obtained by \Eref{Mgut}, as a function of $n$ for
$K=K_1$ (dashed lines) or $K=K_2$ and $K_3$ (solid lines). The
values of the curves which are preferred by the observational data
at $68\%$ c.l. [$95\%$ c.l.] are included in the dark [light] gray
segments -- cf. \Fref{fig1}. We observe that for tiny $n$ values,
$\rs$ which is roughly $1/N$ lies close to $1/3$ for $K=K_1$ and
$1/2$ for $K=K_2, K_3$. For larger $|n|$ values $\rs$ deviates
more drastically from these values.

\begin{figure}[!t]
\centering
\includegraphics[width=60mm,angle=-90]{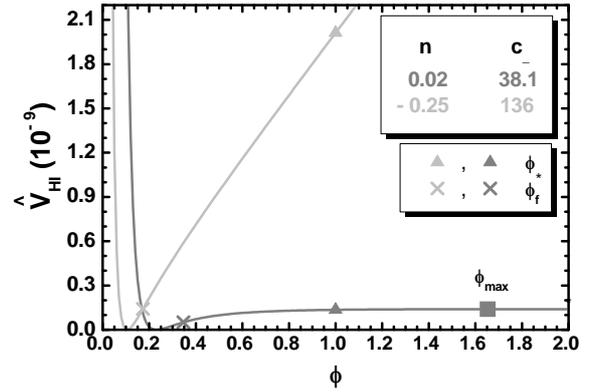}\hspace{-.1mm}
\caption{\sl \small The inflationary potential $\Vhi$ as a
function of $\sg$ for $K=K_2$ or $K_3$, $\sgx=1$ and $n=0.02$
(gray line) or $n=-0.25$ (light gray line). The values
corresponding to $\sgx$, $\sgf$ and $\sg_{\rm max}$ (for $n=0.02$)
are also indicated.}\label{fig2}
\end{figure}

The rather different predictions attained for low ($|n|\leq0.1$)
and large ($|n|>0.1$) $n$ values hint that the structure of $\Vhi$
changes drastically. To illuminate this fact we show  $\Vhi$ as a
function of $\sg$ in \Fref{fig2} for $K=K_2$ or $K_3$, $n=0.02$
(gray line) and $n=-0.25$ (light gray line). We take in both cases
$\sgx=1$. Therefore, the corresponding $\cm$ and $\ld$ values are
confined to their lowest possible values enforcing
\eqs{Ntot}{prob}. More specifically, we find $\ld=9.7\cdot10^{-4}$
or $5.3\cdot10^{-3}$ and $\cm=38.1$ or $136$, with $M=0.23$ or
$M=0.105$ for $n=0.02$ or $n=-0.25$ respectively. The
corresponding $\rs$ values and observable quantities are listed in
\Tref{tab5}. We see that in both cases $\Vhi$ develops a
singularity at $\phi=0$ contrary to the models of non-minimal
inflation -- cf. \cref{univ,jhep} -- where $\Vhi$ exhibits a
maximum. However, for $n>0$ and $|n|\sim0.01$, $\Vhi$ resembles
the potential of Starobinsky model with a maximum at $\sg_{\rm
max}=1.65$. This does not affect much the inflationary evolution
since we find $\Dex=39\%$, and so the tuning of the initial
conditions of IGHI is rather mild. On the contrary, $\Vhi$
increases monotonically and almost linearly with $\sg$ for
$n=-0.25$. Both behaviors can be interpreted from \Eref{Vhi}
taking into account that $\fw\sim\sg^2$ and $\fr\sim\sg^2$. For
$n\sim0.01$, $\Vhi\sim\fw^2/\fr^2$ becomes more or less constant,
whereas for $n\simeq-0.25$,
$\Vhi\sim\fw^2/\fr^{2\cdot3/4}\sim\sg^4/\sg^3\sim\sg$. It is also
remarkable that in the latter case $r$ increases, thanks to the
increase of the inflationary scale, $\Vhi^{1/4}$. Similar region
of parameters is recently reported in \cref{linear}.

\section{A Possible Post-Inflationary Completion} \label{post}

Our discussion about IGHI is certainly incomplete without at least
mentioning how the transition to the radiation dominated era is
realized and the observed BAU is generated. Since these goals are
related to the possible decay channels of the inflaton, the
connection of IGHI with some low energy theory is unavoidable. A
natural, popular and well motivated framework for particle physics
at the $\TeV$ scale beyond the standard model is MSSM. A possible
route to such a more complete scenario is described in
\Sref{post0}. Next, Sec.~\ref{post1} is devoted to the connection
of IGHI with MSSM through the generation of the $\mu$ term. In
Sec.~\ref{post2}, finally, we analyze the scenario of nTL
exhibiting the relevant constraints and further restricting the
parameters. Here and hereafter we restore units, i.e., we take
$\mP=2.433\cdot10^{18}~\GeV$.

\subsection{\sf\scshape\small The Relevant Set-Up} \label{post0}

\renewcommand{\arraystretch}{1.45}
\begin{table*}[!t]
\caption{\normalfont Mass-squared spectrum of the non-inflaton
sector for $K=K_{1\cal R}$ and $K_{2\cal R}$ along the path in
\eqs{inftr}{inftr1}.}
\begin{ruledtabular}
\begin{tabular}{c|c|c|c|c}
{\sc Fields}&{\sc Eigen-} & \multicolumn{3}{c}{\sc Masses
Squared}\\\cline{3-5}
&{\sc states} && {\hspace*{2cm}$K=K_{1\cal R}$\hspace*{2cm}}&{$K=K_{2\cal R}$} \\
\colrule
10 Real & $\widehat{h}_{\pm},\widehat{\bar h}_{\pm}$ &  $ \widehat
m_{h\pm}^2$&$3\Hhi^2\ca\lf
\sg^2/{2N}\pm{2\lm/\ld}\rg$&{$3\Hhi^2\lf1+1/\nsu\pm{4\lm}/{\ld\sg^2}\rg$}\\\cline{3-5}
Scalars  & $\widehat{\tilde\nu}^c_{i},
\widehat{\bar{\tilde\nu}}^c_{i}$ & $\widehat m_{i\tilde
\nu^c}^2$&$3\Hhi^2\ca\lf
\sg^2/2N+8\ld^2_{iN^c}/{\ld^2}\rg$&{$3\Hhi^2\lf1+1/\nsu+16\ld^2_{iN^c
}/\ld^2\sg^2\rg$}\\\colrule
$3$ Weyl Spinors &${\widehat N_i^c}$& $ \widehat
m_{{iN^c}}^2$&\multicolumn{2}{c}{$48\Hhi^2\ld^2_{iN^c
}/\ld^2\sg^2$}
\end{tabular}\label{tab2r}
\end{ruledtabular}
\end{table*}
\renewcommand{\arraystretch}{1.}

\renewcommand{\arraystretch}{1.4}
\begin{table*}[!t]
\caption{\normalfont The same as \Tref{tab2r} but for $K=K_{1},
K_3$ and $K_{3}$.}
\begin{ruledtabular}
\begin{tabular}{c|c|c|c|c|c}
{\sc Fields}&{\sc Eigen-} & \multicolumn{4}{c}{\sc Masses
Squared}\\\cline{3-6}
&{\sc states}&
&{$K=K_1$}&{\hspace*{0.9cm}$K=K_2$\hspace*{0.9cm}} &{$K=K_{3}$}
\\\colrule
10 Real & $\widehat h_{\pm},\widehat{\bar h}_{\pm}$ &  $ \widehat
m_{h\pm}^2$&$3\Hhi^2\lf
1+\fw/{N}\pm{4\lm\cp^2\sg^2}/{\ld\fw}\rg$&\multicolumn{2}{c}{$3\Hhi^2\lf1+1/\nsu\pm{4\lm\cp}/{\ld\fw}\rg$}\\\cline{3-6}
Scalars  & $\widehat{\tilde\nu}^c_{i},
\widehat{\bar{\tilde\nu}}^c_{i}$ &  $ \widehat m_{i\tilde
\nu^c}^2$&$3\Hhi^2\lf
(1+\fw/N)/{\cp\sg^2}+16\ld^2_{iN^c}\cp^2{\sg^2}/{\ld^2\fw^2}\rg$\hspace*{0.3cm}&\multicolumn{2}{c}{$3\Hhi^2\lf1+1/\nsu+16\ld^2_{iN^c
}\cp^2\sg^2/\ld^2\fw^2\rg$}\\\colrule
$3$ Weyl Spinors &${\widehat N_i^c}$& $
m_{{iN^c}}^2$&\multicolumn{3}{c}{$48\ld^2_{iN^c
}\cp^2\sg^2\Hhi^2/\ld^2\fw^2$}
\end{tabular}\label{tab2}
\end{ruledtabular}
\end{table*}
\renewcommand{\arraystretch}{1.}

Following the post-inflationary setting of \cref{univ} we
supplement the superpotential of the theory with the terms
\beqs \beq \Wrhn=\lrh[i]\phcb N^{c2}_i +h_{Nij} \sni L_j \hu,
\label{Wrhn}\eeq
which allows for the implementation of type I see-saw mechanism
(providing masses to light neutrinos) and supports a robust
baryogenesis scenario through nTL, and
\beq\label{Wmu} \Wmu=\lm S\hu\hd\,, \eeq\eeqs
inspired by \cref{dvali}, which offers a solution to one of the
most tantalizing problems of MSSM, namely the generation of a
$\mu$ term -- for an alternative solution see \cref{rsym}. Here we
adopt the notation and the $B-L$ and R charges of the various
superfields as displayed in Table~1 of \cref{univ}. Let us only
note that $L_i$ denotes the $i$-th generation $SU(2)_{\rm L}$
doublet left-handed lepton superfields, and $\hu$ [$\hd$] is the
$SU(2)_{\rm L}$ doublet Higgs superfield which couples to the up
[down] quark superfields. Also, we assume that the superfields
$N_j^c$ have been rotated in the family space so that the coupling
constants $\ld_i$ are real and positive. This is the so-called
\cite{R2r,univ} $\sni$ basis, where the $\sni$ masses, $\mrh[i]$,
are diagonal, real, and positive.

We assume that the extra fields $X^\beta=\hu,\hd,\ssni$ have
identical kinetic terms as the stabilizer field $S$ expressed by
the functions $F_{lS}$ with $l=1,2,3$ in Eqs.~(\ref{f1s}) --
(\ref{f3s}) -- see \cref{univ}. Therefore, $N_S$ may be renamed
$N_X$ henceforth. The inflationary trajectory in \Eref{inftr} has
to be supplemented by the conditions
\beq\label{inftr1} \hu=\hd=\ssni=0,\eeq
and the stability of this path has to be checked, parameterizing
the complex fields above as we do for $S$ in \Eref{hpar}. The
relevant masses squared are listed in \Tref{tab2r} for $K=\kar$
and $\kbr$ and \Tref{tab2} for $K=K_1, K_2$ and $K_3$, where we
see that $\what m^2_{i\tilde\nu^c}>0$ and $\what m^2_{h+}>0$ for
every $\sg$. On the other hand, the positivity of the eigenvalues
$\what m^2_{h-}$ associated with the eigenstates $\widehat h_-$
and $\widehat {\bar h}_-$, where
\beq \widehat h_\pm=(\widehat h_u\pm{\widehat
h_d})/\sqrt{2}\>\>\>\mbox{and}\>\>\> \widehat{\bar
h}_\pm=(\widehat{\bar h}_u\pm\widehat{\bar h}_d)/\sqrt{2}, \eeq
with the hatted fields being defined as $\widehat s$ and
$\widehat{\bar{s}}$ in \Eref{Je1}, requires the establishment of
the inequalities \beqs\bea &\label{lm1r}\hspace*{-8mm}
\lm\lesssim\ld\sg^2/4N
&~\mbox{for}~~K=\kar,\\
 &
\hspace*{-8mm}\label{lm2r}\lm\lesssim\ld\sg^2(1+1/\nsu)/4&~\mbox{for}~~K=\kbr,\\
&\label{lm1}\hspace*{-8mm}
\lm\lesssim\ld\fw(1+\fw/N)/4\lm\cp^2\sg^2
&~\mbox{for}~~K=K_1,\\
 &
\hspace*{-7mm}\label{lm2}\lm\lesssim\ld\fw(1+1/\nsu)/4\cp&~\mbox{for}~~K=K_2,
K_3.~~\eea\eeqs
In all cases, the inequalities are fulfilled for
$\lm\lesssim2\cdot10^{-5}$. Similar numbers are obtained in
\cref{R2r,univ}. We do not consider such a condition on $\lm$ as
unnatural, given that the Yukawa coupling constant $h_{1U}$, which
provides masses to the up-type quarks, is of the same order of
magnitude too at a high scale -- cf. \cref{fermionM}. Note that
the hierarchy in \Erefs{lm1r}{lm2} between $\lm$ and $\ld$ differs
from that imposed in the models \cite{dvali} of F-term hybrid
inflation, where $S$ plays the role of inflaton and
$\phc,\phcb,~\hu$ and $\hd$ are confined at zero. Indeed, in that
case we demand \cite{dvali} $\lm>\ld$ so that the tachyonic
instability in the $\phc-\phcb$ direction occurs first, and the
$\phc-\phcb$ system start evolving towards its v.e.v, whereas
$\hu$ and $\hd$ continue to be confined to zero. In our case,
though, the inflaton is included in the $\bar\Phi-\Phi$ system
while $S$ and the $\hu-\hd$ system are safely stabilized at the
origin both during and after IGHI. Therefore, $\phi$ settles in
its vacuum and $S$, $\hu$ and $\hd$ take their non-vanishing
electroweak scale v.e.vs afterwards.

\subsection{\sf\scshape\small  A Solution to the {$\mu$} Problem of MSSM}
\label{post1}

A byproduct of the $R$ symmetry associated with our models is that
it assists us to understand the origin of the $\mu$ term of MSSM
-- see \Sref{secmu1} -- connecting thereby the high with the low
energy phenomenology as described in \Sref{secmu2}.

\subsubsection{\sf\small Generating the $\mu$
Parameter}\label{secmu1}

Working along the lines of \Sref{hivev} we can verify that the
presence of the terms in \eqs{Wrhn}{Wmu} leave the v.e.vs in
\Eref{vevs} unaltered whereas those of $X^\bt$ are found to be
\beq \vev{\hu}=\vev{\hd}=\vev{\ssni}=0\,.\label{vevs1}\eeq
On the other hand, the contributions from the soft SUSY breaking
terms, although negligible during IGHI -- since these are expected
to be much smaller than $\sg$ --, may slightly shift
\cite{dvali,R2r,univ} $\vev{S}$ from zero in \Eref{vevs}. Indeed,
the relevant potential terms are
\beq V_{\rm soft}= \lf\ld A_\ld S \phcb\phc- {\rm a}_{S}S\ld M^2/4
+ {\rm h. c.}\rg+ m_{\gamma}^2\left|X^\gamma\right|^2,
\label{Vsoft} \eeq
where $X^\gamma=\phc,\phcb,S,\hu,\hd,\ssni$, and $m_{\gamma},
A_\ld$ and $\aS$ are soft SUSY breaking mass parameters of the
order of $\TeV$. The emergence of these terms depend on the
mechanism of SUSY breaking which is not specified here. We
restrict ourselves to the assumption that this extra sector of the
theory may be included in the present set-up without disturbing
the status of inflation -- cf.~\cref{buch}. Confining $\phc,
\phcb, \hu, \hd$ and $\sni$ in their v.e.vs in \eqs{vevs}{vevs1},
$\Ve$ in \Eref{Vsugra} reduces again to $V_{\rm eff}$ in
\Eref{Vsusy} with vanishing terms represented by ellipsis since
$\vev{\Whi}=0$. We then rotate $S$ to the real axis by an
appropriate $R$-transformation and choose conveniently the phases
of $\Ald$ and $\aS$ so as the total low-energy potential
\beq V_{\rm tot}=V_{\rm eff}+V_{\rm soft}\label{vtot}\eeq
to be minimized. Since the form of $V_{\rm eff}$ depends on the
adopted $K$, we single out the cases:

\begin{figure*}[!t]\vspace*{-.12in}
\centering
\includegraphics[width=60mm,angle=-90]{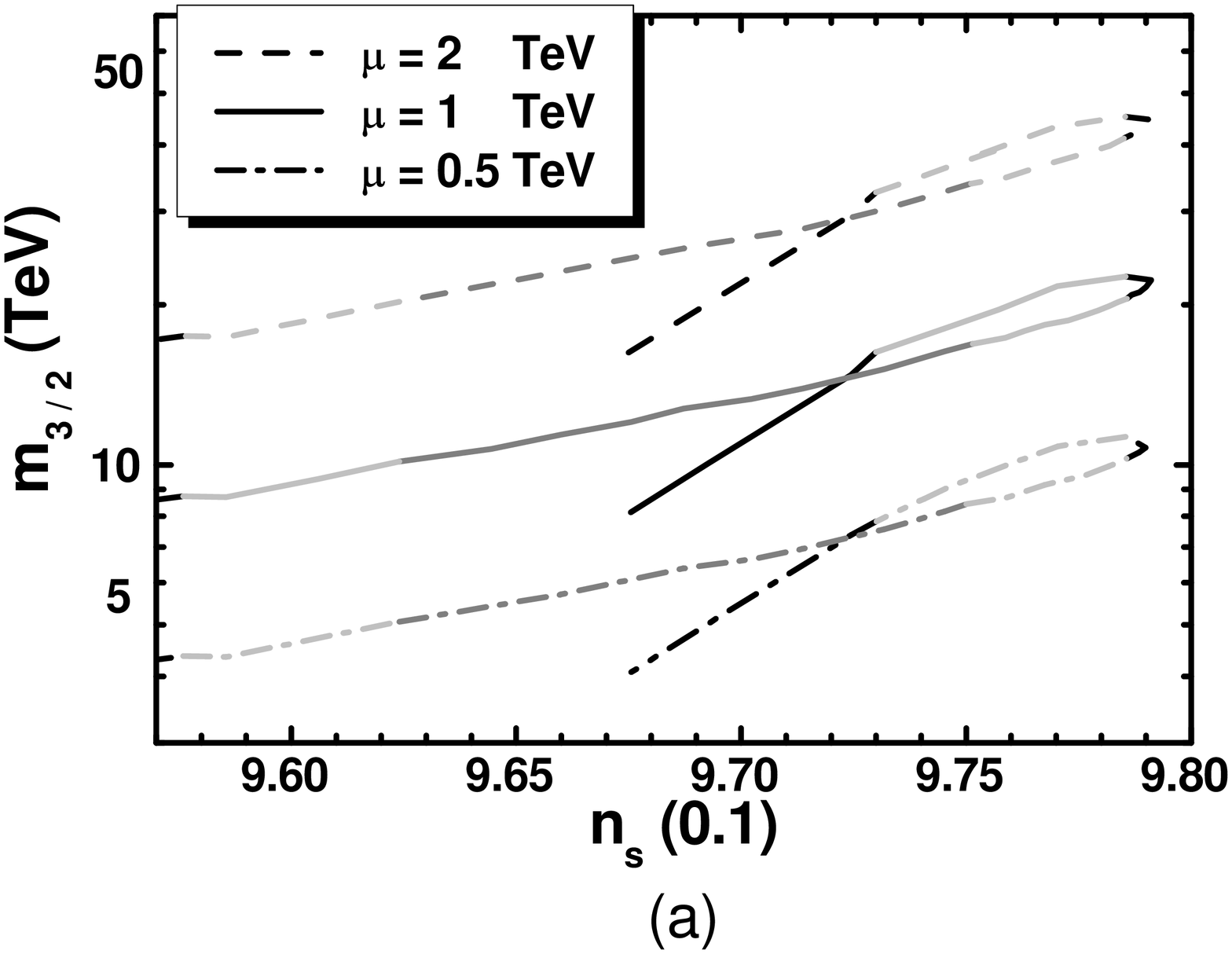}
\hspace{-.1mm}
\includegraphics[width=60mm,angle=-90]{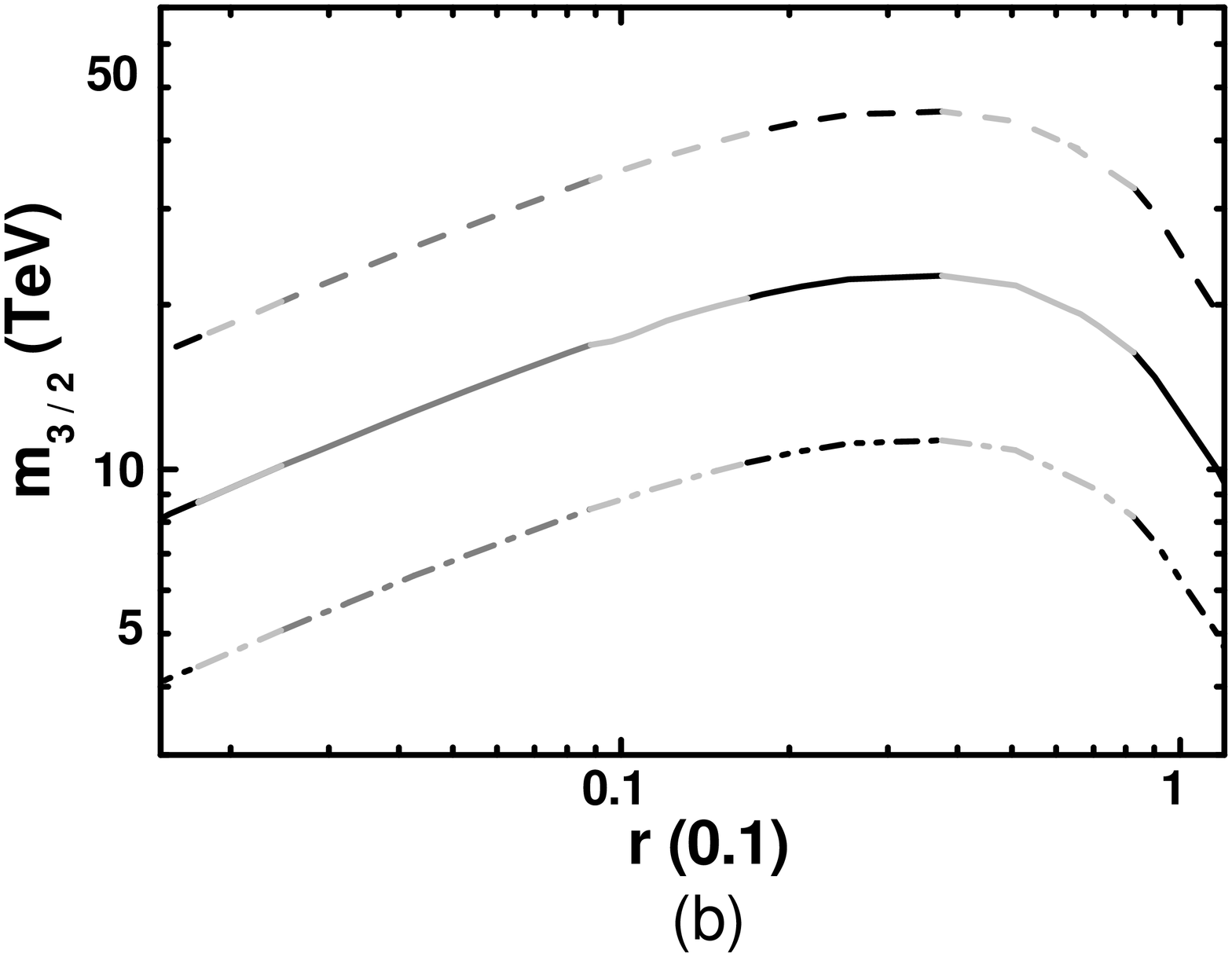}
\caption{\sl\small The gravitino mass $\mgr$ versus $\ns$
{\sffamily\ftn (a)} and $r$ {\sffamily\ftn (b)} for $\lm=10^{-6}$,
$\am=1$,  $K=K_2$ or $K_3$ with $\nsu=2$ and $\mu=0.5~\TeV$
(dot-dashed line), $\mu=1~\TeV$ (solid line), or $\mu=2~\TeV$
(dashed line). The color coding is as in \Fref{nrs}.}\label{fig3}
\end{figure*}

\subparagraph{\sl (a) $K=K_1, K_2$ and $K_3$.} Focusing on $K=K_2$
or $K_3$ we obtain
\beqs\beq \vev{V_{\rm tot}(S)}\simeq\frac{\ld^2S^2}{2\vev{\det
M_\pm}} \lf\cm M^2-N\mP^2\rg -\ld\am \mgr M^2 S, \label{Vol} \eeq
where the first term in the r.h.s originates from the second line
of \Eref{VF} for $e^{\widetilde K_+/\mP^2}\simeq1$, and $\phc$ and
$\phcb$ equal to their v.e.vs in \Eref{vevs}. Also, we take into
account that $m_S\ll M$, and we set
\beq |A_\ld| + |{\rm a}_{S}|=2\am\mgr,\label{am}\eeq
where $\mgr$ is the $\Gr$ mass and $\am>0$ a parameter of order
unity which parameterizes our ignorance of the dependence of
$|A_\ld|$ and $|{\rm a}_{S}|$ on $\mgr$. The minimization
condition for the total potential in \Eref{Vol} w.r.t $S$ leads to
a non vanishing $\vev{S}$ as follows
\beq \label{vevS}\frac{d}{d S} \vev{V_{\rm
tot}(S)}=0~~\Rightarrow~~\vev{S}\simeq \am\mgr\cm(1+N\rs)/\ld,\eeq
since from \eqss{detMpm}{vevs}{ig} we infer
\beq \label{vevdet}\vev{\det
M_\pm}=\cm^2-N^2\cp^2=\cm^2(1+N\rs)(1-N\rs)\,.\eeq
At this $S$ value, $\vev{V_{\rm tot}(S)}$ develops a minimum since
\beq \label{Vss}\frac{d^2}{d S^2} \vev{V_{\rm
tot}(S)}=\ld^2/\cp(\cm+N\cp)>0\,.\eeq\eeqs
For $K=K_1$ \Eref{vevS} can be obtained again by doing an
expansion of the relevant expressions in powers $1/\mP$.

The $\mu$ term generated from \Eref{Wmu} exhibits the mixing
parameter
\beqs\beq\mu =\lm \vev{S}
\simeq\lm\am\mgr\cm(1+N\rs)/\ld\,.\label{mu}\eeq
Comparing this result with the corresponding one in \cref{univ},
we deduce a crucial difference regarding the sign of the
expression in the parenthesis which originates from the terms in
the second line of \Eref{VF}. With the aid of \Eref{lan} we may
eliminate $\cm$ and $\ld$ from the above result which then reads
\beq\mu\simeq1.2\cdot10^{2}
\lm\frac{\am\mgr(1+N\rs)(\frs-1)^{2}}{\rs^{3/2}\frs^{n+1/2}\lf
n(1-\frs)+1\rg}\,,\label{mu1}\eeq\eeqs
where \Eref{prob} is employed to obtain the numerical prefactor.
Taking into account \eqs{Mgut}{sgxb}, we infer that the resulting
$\mu$ depends only on $n$ and not on $\ld, \cm$ and $\rs$ -- cf.
\cref{dvali, univ}. For the $\lm$ values allowed by
\eqs{lm1}{lm2}, any $|\mu|$ value is accessible with a mild
hierarchy between $\mgr$ and $\mu$ -- from \Tref{tab2} we see that
both signs of $\lm$ (and so $\mu$) are possible without altering
the stability analysis of the inflationary system. To understand
this, let us first remark that \Eref{ig} implies $\rs\simeq1/N$
and $\frs$ varies from about $12$ to $68$ [$15$ to $119$] for
$K=K_1$ [$K=K_2$ and $K_3$], as $n$ varies in the allowed ranges
of \Erefs{res1}{res22}. A rough estimation gives
$\mu\sim10^2\lm\frs^{3/2}=10^{-1.5}\mgr$ and so we expect that
$\mu$ is about one order of magnitude less than $\mgr$.

\subparagraph{\sl (b) $K=\kar$ and $\kbr$.} In this case, $V_{\rm
tot}$ in \Eref{vtot} with all the fields except $S$ equal to their
v.e.vs in \eqs{vevs}{vevs1} is written as
\beqs\beq \vev{V_{\rm tot}(S)}= \frac{\ld^2\mP^2S^2}{\ca(N\ca-1)}
-\ld\am \mgr M^2 S\,. \label{Volr} \eeq
The minimization of $\vev{V_{\rm tot}(S)}$ w.r.t $S$ leads to a
new non vanishing $\vev{S}$,
\beq \label{vevSr}~\vev{S}\simeq N \am\mgr\ca/\ld,\eeq
where $M$ is replaced by \Eref{ig}. Therefore, the $\mu$ parameter
involved in \Eref{Wmu} is
\beq\mu =\lm \vev{S}= N\lm\am\mgr\ca/\ld\,.\label{mur}\eeq\eeqs
This still depends only $n$ thanks to the condition in \Eref{lanr}
which fixes $\ld/\ca$ as a function of $n$ -- see \Eref{sgxbr}.

\subparagraph{} To highlight further the conclusions above, we can
employ \Eref{mu} to derive the $\mgr$ values required so as to
obtain a specific $\mu$ value. Given that \Eref{mu} depends on
$n$, which crucially influences $\ns$ and $r$, we expect that the
required $\mgr$ is a function of $\ns$ and $r$ as depicted in
\sFref{fig3}{a} and \sFref{fig3}{b} respectively. We take
$\lm=10^{-6}$, in accordance with \eqs{lm1}{lm2}, $\am=1$, $K=K_2$
or $K_3$ with $\nsu=2$ and $\mu=0.5~\TeV$ (dot-dashed line),
$\mu=1~\TeV$ (solid line), or $\mu=2~\TeV$ (dashed line). Varying
$n$ in the allowed range indicated in \sFref{fig1}{a} we obtain
the variation of $\mgr$ solving \Eref{mu} w.r.t $\mgr$. The values
of the curves which are preferred by the observational data at
$68\%$ c.l. [$95\%$ c.l.] are included in the dark [light] gray
segments -- cf. \Fref{fig1}. We see that $\mgr$ increases with
$\mu$ and its lowest value $\mgr\simeq4~\TeV$ is obtained for
$\mu=0.5~\TeV$. As we anticipated above, $\mgr$ is almost one
order of magnitude larger than the corresponding $\mu$. Moreover,
for fixed $\mu$, each curve develops a maximum at $n\simeq-0.15$,
which coincides with the right corner of the curve in \Fref{fig1}.
This behavior deviates a lot from the one found in \cref{univ} and
comes from the different sign in the parenthesis of \Eref{mu}.

\subsubsection{\sf\small Connection with the Parameters of CMSSM}\label{secmu2}

\renewcommand{\arraystretch}{1.25}
\begin{table}[!t]
\caption[]{\sl\small The required $\lm$ values rendering our
models compatible with the best-fit points of the CMSSM as found
in \cref{mssm} with the assumptions in \Eref{mssmpar}.}
\label{tab}
\begin{ruledtabular}
\begin{tabular}{c|c|l|l|l}
{CMSSM}&$A/H$&$\tilde\tau_1-\chi$ &$\tilde t_1-\chi$&$\tilde
\chi^\pm_1-\chi$\\\cline{3-5}
{\sc Region:}&{\sc Funnel}&\multicolumn{3}{c}{\sc
Coannihilation}\\\colrule
$|A_0|~(\TeV)$&$9.924$& $1.227$&$9.965$&$9.2061$\\
$m_0~(\TeV)$&$9.136$&$1.476$&$4.269$&$9.000$\\
$|\mu|~(\TeV)$&$1.409$&$2.62$&$4.073$&$0.983$\\
$\am$&$1.086$& $0.831$&$2.33$&$1.023$\\
\colrule\multicolumn{5}{c}{$\lm (10^{-6})$ for $K=K_{1\cal
R}$}\\\colrule
$n=0$&$0.963$& ${\it 14.48}$&$2.91$&$0.723$\\
\colrule\multicolumn{5}{c}{$\lm (10^{-6})$ for $K=K_{2\cal
R}$}\\\colrule
$n=0$& $1.184$& $17.81$ &$3.41$ & $0.89$\\
\colrule\multicolumn{5}{c}{$\lm (10^{-6})$ for $K=K_1$}\\\colrule
$n=0.02$&$1.409$& ${\it 21.19}$ &$4.063$&$1.059$\\
$n=-0.25$& $1.87$& $28.11$ &$5.39$ & $1.405$\\
\colrule\multicolumn{5}{c}{$\lm (10^{-6})$ for $K=K_2$ and
$K_3$}\\\colrule
$n=0.02$&$1.814$& ${\it 27.28}$&$5.23$&$1.363$\\
$n=-0.25$& $2.784$& ${\it 41.86}$ &$8.025$ & $2.092$
\end{tabular}
\end{ruledtabular}
\end{table}\renewcommand{\arraystretch}{1.}

The SUSY breaking effects, considered in \Eref{Vsoft}, explicitly
break $U(1)_R$ to a subgroup, $\mathbb{Z}_2^{R}$ which can be
identified with a matter parity. Under this discrete symmetry all
the matter (quark and lepton) superfields change sign -- see
Table~1 of \cref{univ}. Since $S$ has the $R$ symmetry of the
total superpotential of the theory, $\vev{S}$ in \Eref{vevS} also
breaks spontaneously $U(1)_R$ to $\mathbb{Z}_2^{R}$. Thanks to
this fact, $\mathbb{Z}_2^{R}$ remains unbroken and so no
disastrous domain walls are formed. Combining $\mathbb{Z}_2^{R}$
with the $\mathbb{Z}_2^{\rm f}$ fermion parity, under which all
fermions change sign, yields the well-known $R$-parity. This
residual symmetry prevents rapid proton decay and guarantees the
stability of the \emph{lightest SUSY particle} ({\sf\ftn LSP}),
providing thereby a well-motivated \emph{cold dark matter}
({\sf\ftn CDM}) candidate.

The candidacy of LSP may be successful, if it generates the
correct CDM abundance \cite{plcp} within a concrete low energy
framework, which in our case is the MSSM or one of its variants --
for an alternative approach within high-scale SUSY see
\cref{adazi}. Here, we adopt the \emph{Constrained MSSM} ({\ftn\sf
CMSSM}) which is the most restrictive, predictive and
well-motivated version of MSSM, which allows the lightest
neutralino to play the role of LSP in a sizable portion of the
parametric space. This is based on the free parameters
$${\rm sign}\mu,~~\tan\beta=\vev{\hu}/\vev{\hd},~~\mg,~~m_0,~~\mbox{and}~~A_0,$$
where ${\rm sign}\mu$ is the sign of $\mu$, and the three last
mass parameters denote the common gaugino mass, scalar mass and
trilinear coupling constant, respectively, defined (normally) at
$\Mgut$. The parameter $|\mu|$ is not free, since it is computed
at low scale by enforcing the conditions for the electroweak
symmetry breaking. The values of these parameters can be tightly
restricted imposing a number of cosmo-phenomenological constraints
from which the consistency of LSP relic density with observations
plays a central role. Some updated results are recently presented
in \cref{mssm}, where we can also find the best-fit values of
$|A_0|$, $m_0$ and $|\mu|$ listed in the first four lines of
\Tref{tab}.  We see that there are four allowed regions
characterized by the mechanism applied for accommodating an
acceptable CDM abundance.

Taking advantage of this investigation, we can check whether the
$\mu$ and $\mgr$ values satisfying \Eref{mu} are consistent with
these values. Selecting some representative $n$ values and
adopting the identifications
\beq
m_0=\mgr\>\>\>\mbox{and}\>\>\>|A_\ld|=|\aS|=|A_0|,\label{mssmpar}\eeq
we can first derive $\am$ from \Eref{am} and then the $\lm$ values
from \Erefs{mu}{mur}, which yield the phenomenologically desired
$|\mu|$ shown in the third line of \Tref{tab}. Here we assume that
renormalization effects in the derivation of $\mu$ are negligible.
The outputs of our computation are assembled in the last ten lines
of \Tref{tab}. As inputs, we take $n=0.0$ for $K=\kar$ and $\kbr$
$n=0.02$ and $-0.25$ for $K=K_1, K_2$ and $K_3$. These are central
values in the regions compatible with the inflationary
observations as found in \Sref{hinum}. The $\lm$ values for
$K=\kar$ and $\kbr$ are lower than those obtained for $K=K_1, K_2$
and $K_3$, larger than those found in \cref{univ}, and similar to
those in \cref{R2r} especially for $K=K_{1\cal R}$. On the other
hand, the $\lm$ values found for $K=K_1, K_2$ and $K_3$ are larger
compared to those found in \crefs{R2r,univ}.

From the outputs we infer that the required $\lm$ values are
comfortably compatible with Eqs. (\ref{lm1r}) -- (\ref{lm2}) for
$\nsu=2$, in all cases besides the one corresponding to the
$\tilde\tau_1-\chi$ coannihilation region. In that case, $m_0$ is
lower than $|\mu|$ and so marginally large $\lm$ values are
required. In the cases where numbers are written in italics, we
obtain instability along the inflationary path for $K=\kar$ and
$K_1$, whereas for $K=\kbr$, $K_2$ and $K_3$ we need $0\leq
N_X\leq1$ to avoid this effect. In sharp contrast to the model of
\cref{univ}, only the $A/H$ funnel and $\tilde\chi^\pm_1-\chi$
coannihilation regions can be consistent with the $\Gr$ limit on
$\Trh$ -- see \Sref{lept2}. Indeed, $\mgr\gtrsim9~\TeV$ become
cosmologically safe under the assumption of an unstable $\Gr$, for
the $\Trh$ values necessitated for satisfactory leptogenesis --
see \Sref{leptres}.

\subsection{\sf\scshape\small  Non-Thermal Leptogenesis}\label{post2}

Our next task is to specify how our inflationary scenario makes a
transition to the radiation dominated era -- see \Sref{lept1} --
and offers an explanation of the observed BAU consistent with the
$\Gr$ constraint and the low energy neutrino data -- see
\Sref{lept2}. Our results are summarized in \Sref{leptres}.

\subsubsection{\small\sf Inflaton Mass and Decay}\label{lept1}

When IGHI is over, the inflaton continues to roll down towards the
SUSY vacuum, \Eref{vevs}. Soon after, it settles into a phase of
damped oscillations around the minimum of $\Vhi$. The (canonically
normalized) inflaton,
\beq\dphi=\vev{J}\dph\>\>\>\mbox{with}\>\>\> \dph=\phi-M,
\label{dphi}\eeq and
\beq\vev{J}=\begin{cases}\sqrt{N\ca}&\mbox{for}~~K=\kar~~\mbox{and}~~\kbr,
\\\sqrt{\cm(1+{N\rs})}&\mbox{for}~~K=K_1,K_2~~\mbox{and}~~K_3,\end{cases}\eeq
acquires mass, at the SUSY vacuum in \Eref{vevs}, which is given
by
\beq \label{msn} \frac\msn{\ld
\mP}=\begin{cases}\sqrt{\ca\lf{N\ca}-1\rg}&\mbox{for}~~K=\kar~~\mbox{and}~~\kbr,
\\\cm\sqrt{2\lf1+{N\rs}\rg}&\mbox{for}~~K=K_1,K_2~~\mbox{and}~~K_3.\end{cases}
\eeq
From the last expression we can infer that $\msn$ remains constant
for fixed $n$ since $\ld/\ca$ [$\ld/\cm$] is fixed too -- see
\eqs{lanr}{lan}. More specifically, for the allowed range of $n$
in \eqs{res1r}{res2r} we obtain
\beqs\bea \label{resmass1r} & 2.\lesssim
{\msn/10^{13}~\GeV}\lesssim3.9&\mbox{for}\>\>K=\kar,\\
\label{resmass2r} &2.1\lesssim {\msn/10^{13}~\GeV}\lesssim4.5\,&
\mbox{for}\>\>K=\kbr,~~~~~~~ \eea\eeqs
with the value $\msn=2.8\cdot10^{13}~\GeV$ corresponding to $n=0$.
Furthermore, for $K=K_1$ and the allowed range of $n$ in
\Eref{res1} we obtain
\beqs\beq \label{resmass1} 2.9\lesssim
{\msn/10^{13}~\GeV}\lesssim5.\eeq
For $K=K_2$ and $K_3$, in the allowed ranges of
\eqs{res21}{res22}, we obtain
\bea \label{resmass21} &3.1\lesssim
{\msn/10^{13}~\GeV}\lesssim6.9;\\
&4.1\lesssim {\msn/10^{13}~\GeV}\lesssim7.2\,.
\label{resmass22}\eea\eeqs
We remark that $\msn$ is somewhat affected by the choice of $K$'s
in Eqs.~(\ref{K1}) -- (\ref{K3}). For $n=0$,
$\msn=4.7\cdot10^{13}~\GeV$ for $K=K_1$, and
$\msn=5.2\cdot10^{13}~\GeV$ for $K=K_2$ and $K_3$, which are both
somewhat larger than the value obtained within Starobinsky
inflation \cite{R2r, su11}. On the other hand, these values are
close to the maximal ones found in \cref{univ}, since here $\rs$
approaches its maximal value.

The inflaton can decay \cite{Idecay} perturbatively into:
\subparagraph{\sl (a)} A pair of $N^c_{i}$ with Majorana masses
$\mrh[i]=\ld_{iN^c}M$ with the decay width
\beqs\beq \label{GNN}
\GNsn=\frac{g_{iN^c}^2}{16\pi}\msn\lf1-\frac{4\mrh[
i]^2}{\msn^2}\rg^{3/2},\eeq where the relevant coupling
constant\beq \label{gNN}
g_{iN^c}=(N-1)\frac{\ld_{iN^c}}{\vev{J}}\eeq arises from the
lagrangian term \bea\nonumber{\cal L}_{\dphi\to
\sni\sni}&=&-\frac12e^{K/2\mP^2}W_{{\rm RHN},N_i^cN^c_i}\sni\sni\
+{\rm h.c.}\\ &=&g_{iN^c} \dphi\ \sni\sni\ +{\rm h.c.}
\label{lNN}\eea\eeqs
This decay channel activates the mechanism of nTL, as sketched in
\Sref{lept2}.

\subparagraph{\sl (b)} Higgses $\hu$ and $\hd$ with the decay
width
\beqs\beq \label{Ghh}
\Ghsn=\frac{2}{8\pi}g_{H}^2\msn\,\>\>\mbox{where}\>\>\>g_{H}=\frac{\lm}{\sqrt{2}}
\eeq arises from the lagrangian term \bea \nonumber {\cal
L}_{\dphi\to\hu\hd}&=&-e^{K/\mP^2}K^{SS^*}\left|W_{\mu,S}\right|^2
\\&=&-g_{H} \msn\dphi \lf H_u^*H_d^*+{\rm h.c.}\rg+\cdots
\label{lhh}\eea\eeqs
Thanks to the upper bounds on $\lm$ from \eqs{lm1}{lm2}, $g_{H}$
turns out to be comparable with $g_{iN^c}$.

\subparagraph{\sl (c)} MSSM (s)-particles $XYZ$ with the following
$\cp$-dependent 3-body decay width
\beqs\beq \label{Gxyz} \Gysn=g_y^2\frac{n_{\rm
f}}{512\pi^3}\frac{\msn^3}{\mP^2},\eeq where for the third
generation we take $y\simeq(0.4-0.6)$, computed at the $\msn$
scale, and $n_{\rm f}=14$ for $\msn<\mrh[3]$. Also, \beq
\label{gxyz} g_y=y_{3}\cdot\begin{cases}
\sqrt{({N\ca-1})/{2\ca}}&\mbox{for}~~K=\kar~~\mbox{and}~~\kbr,\\
N\sqrt{{\rs}/({1+N\rs})}&\mbox{for}~~K=K_1,K_2~~\mbox{and}~~K_3\end{cases}\eeq
and $y_{3}=h_{t,b,\tau}(\msn)\simeq0.5\,$.
Since $\rs\simeq1/N$ we can easily infer that $g_y$ above is
enhanced compared to the corresponding one in \cref{univ} where
$\rs\simeq0.01$, and an additional suppression through a ratio
$M/\mP$ exists. We therefore expect that $\Gysn$ contributes
sizably to the total decay width of $\dphi$. Each individual decay
width arises from the langrangian terms \bea \nonumber{\cal
L}_{\dphi\to X\psi_Y\psi_Z}&=&-\frac12e^{{K/2\mP^2}}\lf
W_{y,YZ}\psi_{Y}\psi_{Z}\rg+{\rm h.c.}\,\\
&=&-g_y\frac\dphi\mP\lf X\psi_{Y}\psi_{Z}\rg+{\rm
h.c.},\label{lxyz} \eea\eeqs where $W_y=yXYZ$ is a typical
trilinear superpotential term of MSSM with $y$ a Yukawa coupling
constant, and $\psi_X, \psi_{Y}$ and $\psi_{Z}$ are the chiral
fermions associated with the superfields $X, Y$ and $Z$ whose
scalar components are denoted with the superfield symbols.

\subparagraph{} The resulting reheat temperature is given by
\cite{quin}
\beqs \beq \label{T1rh} \Trh=
\left({72/5\pi^2g_*}\right)^{1/4}\Gsn^{1/2}\mP^{1/2},\eeq with the
total decay width of $\dphi$ being
\beq\Gsn=\GNsn+\Ghsn+\Gysn\,.\label{Trh}\eeq\eeqs
Here, $g_*=228.75$ counts the MSSM effective number of
relativistic degrees of freedom at temperature $T_{\rm rh}$. Let
us clarify here that in our models there is no decay of a scalaron
as in the original (non-SUSY) \cite{R2} Starobinsky inflation and
some \cite{ketov} of its SUGRA realizations; thus, $\Trh$ in our
case is slightly lower than that obtained there.

\subsubsection{\small\sf Lepton-Number and Gravitino Abundances}\label{lept2}

For $\Trh<\mrh[i]$, the out-of-equilibrium decay of  $\nu^c_{i}$
generates a lepton-number asymmetry (per $\nu^c_{i}$ decay),
$\ve_i$ -- see, e.g., \cref{baryo, lept}. The resulting $\ve_i$ is
partially converted through sphaleron effects into a yield of the
observed BAU \cite{baryo, lept, univ},
\beq Y_B=-0.35\cdot2\cdot{5\over4}{\Trh\over\msn}\mbox{$\sum_i$}
{\GNsn\over\Gsn}\ve_i,\label{Yb}\eeq
which has to reproduce the observational result \cite{plcp}
\beq Y_B=\lf8.64^{+0.15}_{-0.16}\rg\cdot10^{-11}.\label{Ybw}\eeq
The validity of \Eref{Yb} requires that the $\dphi$ decay into a
pair of $\sni$'s is kinematically allowed for at least one species
of the $\sni$'s and also that there is no erasure of the produced
$Y_L$ due to $N^c_1$ mediated inverse decays and $\Delta L=1$
scatterings \cite{senoguz}. These prerequisites are ensured if we
impose
\beq\label{kin} {\sf \ftn
(a)}\>\>\msn\geq2\mrh[1]\>\>\>\mbox{and}\>\>\>{\sf \ftn
(b)}\>\>\mrh[1]\gtrsim 10\Trh.\eeq
The quantity $\ve_i$ can be expressed in terms of the Dirac masses
of $\nu_i$, $\mD[i]$, arising from the second term of \Eref{Wrhn}
-- see \cref{univ}. Moreover, employing the seesaw formula we can
then obtain the light-neutrino mass matrix $m_\nu$ in terms of
$\mD[i]$ and $\mrh[i]$. As a consequence, nTL can be nicely linked
to low energy neutrino data. We take as inputs the best-fit values
\cite{valle} -- see also \cref{lisi} -- of the neutrino
mass-squared differences, $\Delta m^2_{21}=7.6\cdot10^{-5}~{\rm
eV}^2$ and $\Delta m^2_{31}=\lf
2.48\left[-2.38\right]\rg\cdot~10^{-3}~{\rm eV}^2$, of the mixing
angles, $\sin^2\theta_{12}=0.323$,
$\sin^2\theta_{13}=0.0226~\left[\sin^2\theta_{13}=0.029\right]$
and
$\sin^2\theta_{23}=0.567~\left[\sin^2\theta_{23}=0.573\right]$,
and of the CP-violating Dirac phase
$\delta=1.41\pi~\left[\delta=1.48\pi\right]$ for \emph{normal
[inverted] ordered} (NO [IO]) \emph{neutrino masses}, $\mn[i]$'s.
The sum of $\mn[i]$'s is bounded from above by the data
\cite{plcp}, \beq \label{summn}\mbox{$\sum_i$}
\mn[i]\leq0.23~{\eV}\eeq at 95\% c.l. This is more restrictive
than the $90\%$ c.l. upper bound arising from the effective
electron neutrino mass in $\beta$-decay \cite{2beta}: \beq
\label{2beta} m_{\beta}\leq(0.061-0.165)~\eV,\eeq where the range
accounts for nuclear matrix element uncertainties.

The required $\Trh$ in \Eref{Yb} must be compatible with
constraints on the $\Gr$ abundance, $Y_{3/2}$, at the onset of
\emph{nucleosynthesis} (BBN), which are \cite{kohri} given
approximately by
\beq  \label{Ygw} Y_{3/2}\lesssim\left\{\bem
%
10^{-14}\hfill \cr
10^{-13}\hfill \cr
10^{-12}\hfill \cr\eem
\right.\>\>\>\mbox{for}\>\>\>m_{3/2}\simeq\left\{\bem
0.69~\TeV,\hfill \cr
10.6~\TeV,\hfill \cr
13.5~\TeV.\hfill \cr\eem
\right.\eeq
Here we consider the conservative case where $\Gr$ decays with a
tiny hadronic branching ratio. The bounds above can be somehow
relaxed in the case of a stable $\Gr$ -- see e.g. \cref{okada}.

In our models $Y_{3/2}$ is estimated to be \cite{brand,kohri}:
\beq\label{Ygr} Y_{3/2}\simeq 1.9\cdot10^{-22}\ \Trh/\GeV ,\eeq
where we take into account only thermal production of $\Gr$, and
assume that $\Gr$ is much heavier than the MSSM gauginos.
Non-thermal contributions to $\Yg$ \cite{Idecay} are also possible
but strongly dependent on the mechanism of soft SUSY breaking.
Moreover, no precise computation of this contribution exists
within IGHI adopting the simplest Polonyi model of SUSY breaking
\cite{grNew}. It is notable, though, that the non-thermal
contribution to $\Yg$ in models with stabilizer field, as in our
case, is significantly suppressed compared to the thermal one.

\subsubsection{\small\sf  Results}\label{leptres}

It is worthwhile to test the applicability of the framework above
in the case of IGHI. Namely, following a bottom-up approach
detailed in \cref{univ}, we find the $\mrh[i]$'s by using as
inputs the $\mD[i]$'s, a reference mass of the $\nu_i$'s --
$\mn[1]$ for NO $\mn[i]$'s, or $\mn[3]$ for IO $\mn[i]$'s --, the
two Majorana phases $\varphi_1$ and $\varphi_2$ of the PMNS
matrix, and the best-fit values, mentioned in \Sref{lept2}, for
the low energy parameters of neutrino physics -- note that there
are no experimental constraints on $\varphi_1$ and $\varphi_2$ up
to now. In our numerical code we also estimate, following
\cref{running}, the renormalization group evolved values of the
latter parameters at the scale of nTL, $\Lambda_L=\msn$, by
considering the MSSM with $\tan\beta\simeq50$ as an effective
theory between $\Lambda_L$ and the soft SUSY breaking scale,
$M_{\rm SUSY}=1.5~\TeV$. We evaluate the $\mrh[i]$'s at
$\Lambda_L$, and we neglect any possible running of the $\mD[i]$'s
and $\mrh[i]$'s. Therefore, we present their values at
$\Lambda_L$.

\renewcommand{\arraystretch}{1.4}
\begin{table}[!t]
\caption{\sl Parameters yielding the correct BAU for $K=K_2$ or
$K_3$, $n=0.02$, $\lm=10^{-6}$, $y_3=0.5$ and various neutrino
mass schemes.}
\begin{ruledtabular}\begin{tabular}{c||c|c||c|c|c||c|c}
{\sc Parameters} &  \multicolumn{7}{c}{\sc Cases}\\\cline{2-8}
&A&B& C & D& E & F&G\\ \cline{2-8} &\multicolumn{2}{c||}{Normal} &
\multicolumn{3}{c||}{Almost}&  \multicolumn{2}{c}{Inverted}
\\& \multicolumn{2}{c||}{Hierarchy}&\multicolumn{3}{c||}{Degeneracy}& \multicolumn{2}{c}{Hierarchy}\\
\colrule
\multicolumn{8}{c}{Low Scale Parameters (Masses in
$\eV$)}\\\colrule
$\mn[1]/0.1$&$0.01$&$0.1$&$0.5$ & $0.7$& $0.7$ & $0.5$&$0.49$\\
$\mn[2]/0.1$&$0.09$&$0.13$&$0.51$ & $1.0$& $0.705$ & $0.51$&$0.5$\\
$\mn[3]/0.1$&$0.5$&$0.51$&$0.71$ & $1.12$&$0.5$ &
$0.1$&$0.05$\\\colrule
$\sum_i\mn[i]/0.1$&$0.6$&$0.74$&$1.7$ & $2.3$&$1.9$ & $1.1$&$1$\\
$m_\beta/0.01$&$0.22$&$0.98$&$3.5$ & $5.3$&$2.9$ &
$4.9$&$3.6$\\
\colrule
$\varphi_1$&$0$&$0$&$0$ & $\pi/2$&$\pi/2$ & $-3\pi/4$&$0$\\
$\varphi_2$&$-\pi/2$&$0$ &$\pi/2$& $-\pi$&$-2\pi/3$ &
$5\pi/4$&$-\pi/2$\\\colrule
\multicolumn{8}{c}{Leptogenesis-Scale Mass Parameters in
$\GeV$}\\\colrule
$\mD[1]$&$1.98$&$1.5$&$2.3$ & $4.16$&$5.2$ & $1$&$6.3$\\
$\mD[2]$&$38$&$16.6$&$12$ & $10$&$9.6$ & $6.6$&$10$\\
$\mD[3]/100$&$1$&$1$&$1$ & $1$&$1$ & $1$&$0.33$\\\colrule
$\mrh[1]/10^{11}$&$1.6$&$2.1$&$1.4$ & $2.8$&$5.2$ & $0.2$&$8.9$\\
$\mrh[2]/10^{12}$& $27$&$6.8$&$2.6$&$4.8$ &$1.9$ & $2.2$&$3.2$\\
$\mrh[3]/10^{14}$&$22$&$4.7$&$0.89$ & $0.22$&$0.69$ &
$2.9$&$0.9$\\\colrule
\multicolumn{8}{c}{Decay channels of the Inflaton
$\dphi$}\\\colrule
$\dphi\to$&$\wrhn[1]$&$\wrhn[1,2]$& $\wrhn[1,2]$& $\wrhn[1,2]$&
$\wrhn[1,2]$ & $\wrhn[1,2]$&$\wrhn[1,2]$\\
\colrule
\multicolumn{8}{c}{Resulting $B$-Yield }\\\colrule
$Y^0_B/10^{-11}$&$9.63$&$8$& $8.4$& $9.1$&$8.9$ & $8.7$&$8.9$\\
$Y_B/10^{-11}$&$8.67$&$8.59$& $8.69$& $8.56$&$8.65$ &
$8.67$&$8.65$\\\colrule
\multicolumn{8}{c}{Resulting $\Trh$ (in $\GeV$) and $\Gr$-Yield
}\\\colrule
$\Trh/10^{9}$&$1$&$1.1$& $1$& $1.1$&$1$ & $1$&$1$\\
$10^{13}Y_{3/2}$&$1.91$&$2.2$& $1.9$& $2$&$1.9$ & $1.9$&$1.97$
\end{tabular}
\end{ruledtabular} \label{tab4}
\end{table}
\renewcommand{\arraystretch}{1.}

We start the exposition of our results arranging in \Trefs{tab4}
for $K=K_2$ or $K_3$ and \ref{tab4r} for $K=\kbr$ some
representative values of the parameters which yield $\Yb$ and
$\Yg$ compatible with \eqs{Ybw}{Ygw}, respectively. Throughout our
computation we take $\lm=10^{-6}$, in accordance with
\eqs{lm1}{lm2}, and $y=0.5$, which is a typical value encountered
\cite{fermionM} in various MSSM settings with large $\tan\beta$.
Also, we select $n=0.02$ in \Tref{tab4} and $n=0$ in \Tref{tab4r}.
These values yield $\ns$ and $r$ in the ``sweet'' spot of the
present data -- see \Frefs{fig1} and \ref{fig1r}. We obtain
$M=2.85 \cdot 10^{16}~\GeV$ and $\msn=2.8\cdot10^{13}~\GeV$ for
$K=\kar$ or $\kbr$, $M=6.1\cdot10^{17}~\GeV$ and
$\msn=4.2\cdot10^{13}~\GeV$ for $K=K_1$, or
$M=5.6\cdot10^{15}~\GeV$ and $\msn=4.4\cdot10^{13}~\GeV$ for
$K=K_2$ or $K_3$. Although the uncertainties from the choice of
$K$'s are negligible as regards the quantities above, the decay
widths in \Sref{lept1} depend on $N$ (and $\rs$) which take
slightly different values for $K=\kar$ or $K_1$ and $K=\kbr, K_2$
or $K_3$ -- see e.g. \Fref{nrs} -- discriminating somehow the
various choices. For this reason, we clarify that we adopt
$K=\kbr$ in \Tref{tab4r} and $K=K_2$ or $K_3$ in \Tref{tab4}. Had
we employed $K=\kar$ or $K_1$, we would have obtained almost two
times larger $\Yb$'s with the same values of the free parameters.
Therefore a mild readjustment is needed.

In both Tables we consider NO (cases A and B), almost degenerate
(cases C, D and E) and IO (cases F and G) $\mn[i]$'s. In all cases
\Eref{summn} is safely met -- the case D saturates it -- whereas
\Eref{2beta} is comfortably satisfied. The gauge group adopted
here, $G_{B-L}$, does not predict any relation between the Yukawa
couplings constants $h_{iN}$ entering the second term of
\Eref{Wrhn} and the other Yukawa couplings in the MSSM. As a
consequence, the $\mD[i]$'s are free parameters. However, for the
sake of comparison, for cases A -- F, we take
$\mD[3]=m_t(\Lambda_L)\simeq100~\GeV$, where $m_t$ denotes the
mass of the top quark. Similar conditions for the lighter
generations do not hold, though, in our data sample.

\renewcommand{\arraystretch}{1.4}
\begin{table}[!t]
\caption{\sl Same as in \Tref{tab4} but for $K=\kbr$ and $n=0$.}
\begin{ruledtabular}
\begin{tabular}{c||c|c||c|c|c||c|c}
{\sc Parameters} &  \multicolumn{7}{c}{\sc Cases}\\\cline{2-8}
&A&B& C & D& E & F&G\\ \colrule
\multicolumn{8}{c}{Low Scale Parameters as in
\Tref{tab4}}\\\colrule
\multicolumn{8}{c}{Leptogenesis-Scale Mass Parameters in
$\GeV$}\\\colrule
$\mD[i]^{(*)}$&$1.91$&$16.6$&$11.6$ & $10.15$&$9.25$ &
$6.37$&$9.5$\\\colrule
\multicolumn{8}{c}{$^{(*)}$ Where $i=1$ for case A and $i=2$ for
the others.}\\
\multicolumn{8}{c}{The remaining $\mD[i]$ and $\mrh[i]$ are as in
\Tref{tab4}.}\\\colrule
\multicolumn{8}{c}{Resulting $B$-Yield }\\\colrule
$Y^0_B/10^{-11}$&$9.6$&$7.8$& $8.5$& $8.9$&$8.9$ & $8.9$&$8.7$\\
$Y_B/10^{-11}$&$8.64$&$8.61$& $8.72$& $8.6$&$8.73$ &
$8.8$&$8.5$\\\colrule
\multicolumn{8}{c}{Resulting $\Trh$ (in $\GeV$) and $\Gr$-Yield
}\\\colrule
$\Trh/10^{8}$&$7.6$&$8.4$& $7.7$& $8.4$&$7.6$ & $7.6$&$7.8$\\
$Y_{3/2}/10^{-13}$&$1.44$&$2.9$& $1.5$& $1.6$&$1.45$ &
$1.45$&$1.5$
\end{tabular}
\end{ruledtabular} \label{tab4r}
\end{table}
\renewcommand{\arraystretch}{1.}


Besides case A, where only the channel $\dphi\to N^c_1N^c_1$ is
kinematically unblocked, $\dphi$ decays into $N_1^c$'s and
$N_2^c$'s. In the latter cases $\ve_2$ yields the dominant
contribution to the calculation $\Yb$ from \Eref{Yb}. From our
computation, we also remark that $\GNsn<\Ghsn<\Gysn$, and so the
ratios $\GNsn/\Gsn$ introduce a considerable reduction
($0.02-0.25$) in the derivation of $\Yb$. As a consequence, the
attainment of the correct $\Yb$ requires relatively large
$\mD[i]$'s with $i=1,2$ in order to achieve sizable enough
$\GNsn$. Namely, $\mD[1]\gtrsim1~\GeV$ and
$\mD[2]\gtrsim6.6~\GeV$. Besides case A, the first inequality is
necessary, in order to fulfill the second inequality in
\Eref{kin}, given that $\mD[1]$ heavily influences $\mrh[1]$. In
\Tref{tab4r} we list only $\mD[1]$ in case A or $\mD[2]$ in the
other cases which are adjusted so as to accommodate $\Yb$ within
the range of \Eref{Ybw} with the others $\mD[i]$ remaining as
shown in \Tref{tab4}. As a consequence, $\mrh[i]$ deviate very
little from the values shown in \Tref{tab4}.

In both Tables we also display, for comparison, the $B$-yield with
($Y_B$) or without ($Y^0_B$) taking into account the
renormalization group effects. We observe that the two results are
mostly close to each other. Shown also are values for $\Trh$, the
majority of which are close to $10^9~\GeV$, and the corresponding
$Y_{3/2}$'s, with the results for $K=\kbr$ being a little lower.
Thanks to our non-thermal set-up, successful leptogenesis can be
accommodated with $\Trh$'s lower than those necessitated in the
thermal regime -- cf. \cref{stefan}. The resulting large
$Y_{3/2}$'s may be consistent with \Eref{Ygw} mostly for
$m_{3/2}\gtrsim10~\TeV$. These are marginally tolerated with the
$\mgr$'s appearing in \Tref{tab} and Figs.~2 and 3 of \cref{mssm}
in the $A/H$ funnel and $\chi_1^\pm-\chi$ coannihilation regions
-- see also \cref{nick}. These $\mgr$'s though are more easily
reconciled with low energy data in less restrictive  versions of
MSSM -- see e.g. \cref{baer}.

\begin{figure}[!t]
\centering\includegraphics[width=60mm,angle=-90]{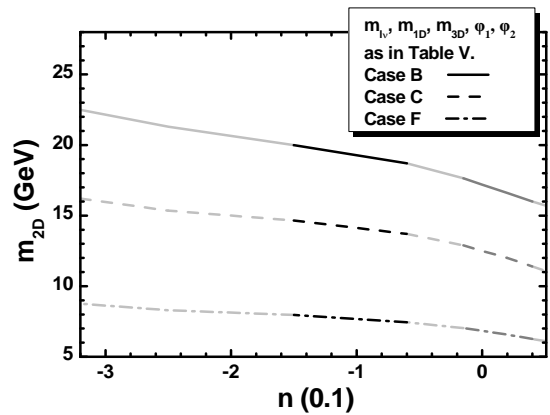}
\caption{\sl\small Contours in the $n-m_{\rm 2D}$ plane yielding
the central $Y_B$ in \Eref{Ybw} consistently with the inflationary
requirements for $K=K_2$ or $K_3$, $\lm=10^{-6}$, $y_3=0.5$ and
the values of $m_{i\nu}$, $m_{\rm 1D}$, $m_{\rm 3D}$, $\varphi_1$,
and $\varphi_2$ which correspond to the cases B (solid line), C
(dashed line), and F (dot-dashed line) of \Tref{tab4}. The color
coding is as in \Fref{nrs}.}\label{nm2D}
\end{figure}

In order to extend the conclusions inferred from \Tref{tab4} to
the case of variable $n$, we can examine how the central value of
$Y_B$ in \Eref{Ybw} can be achieved by varying $m_{\rm 2D}$ as a
function of $n$. The resulting contours in the $n-m_{\rm 2D}$
plane are presented in \Fref{nm2D} -- since the range of $Y_B$ in
\Eref{Ybw} is very narrow, the $95\%$ c.l. width of these contours
is negligible. The convention adopted for these lines is also
described in the figure. In particular, we use solid, dashed, or
dot-dashed line for $\mn[i]$, $\mD[1]$, $\mD[3]$, $\varphi_1$, and
$\varphi_2$ corresponding to the cases B, D, or F of \Tref{tab4}
respectively. For $n$ within its allowed margins in
\eqs{res21}{res22} we obtain
$0.4\lesssim\Trh/10^9~\GeV\lesssim1.8$, which is perfectly
acceptable from \Eref{Ygw} for $\mgr\gtrsim10~\TeV$. Along the
depicted contours, the resulting $\mrh[2]$'s vary in the ranges
$(5.7-14.5)\cdot10^{12}~\GeV$, $(1.8-4.6)\cdot10^{12}~\GeV$,
$(1.5-3.9)\cdot10^{12}~\GeV$  for cases B, C and F respectively,
whereas $\mrh[1]$ and $\mrh[3]$ remain close to their values
presented in the corresponding cases of \Tref{tab4}.

Comparing, finally, our results above with those presented in
\cref{univ}, we can deduce that here $\msn$ and $\Trh$ gain almost
their maximal allowed values since $\rs$ is also maximized due to
the hypothesis of \Eref{Mgut}. As a consequence, $\mgr$ also has
to be enhanced to avoid problems with BBN, whereas $\mD[1,2]$ and
$\mrh[1,2]$ are also constrained to larger values. On the other
hand, our results are closer to those obtained employing the model
of IG in \cref{R2r} with gauge singlet inflaton and without
unification constraint.

\section{Conclusions}\label{con}

We have proposed a class of novel inflationary models, in which a
Higgs field plays the role of the inflaton, before settling in its
final vacuum state where it generates the Planck scale  and gives
rise to a mass for the gauge boson consistent with gauge coupling
unification within MSSM. These two hypotheses allow us to
determine the mass scale $M$, entering $\Whi$ in Eq.~(\ref{Whi}),
and $\ca$ for the $K$'s in \eqs{K1r}{K2r} or $\rs=\cp/\cm$ for the
$K$'s in \Erefs{K1}{K3}. In the latter cases, $\rs$ expresses the
amount of violation of a shift symmetry. As a consequence, the
inflationary scenario depends essentially on two free parameters
-- $n$ and $\ld$ or $\ld/\cm$ for the first or second  group of
$K$'s, respectively -- leading naturally to observationally
acceptable results. Namely, for the $K$'s in \Erefs{K1}{K3} we
obtained slightly larger $r$'s and two distinct allowed regions of
parameters with $n$ values one order of magnitude larger than
those needed for the $K$'s in \eqs{K1r}{K2r}. As an example, the
model for $K=K_2$ or $K_3$, $n=0$ and $\ld/\cm=3\cdot10^{-5}$
yields $\ns\simeq0.973$ and $r\simeq0.0066$ with negligibly small
$\as$. In all cases, inflation is attained for \sub\ inflaton
values, thereby stabilizing our predictions from possible higher
order corrections, whereas the corresponding effective theories
remain trustable up to $\mP$. 

The models were further extended to generate the MSSM $\mu$
parameter, consistently with the low energy phenomenology.
Successful baryogenesis is achieved via primordial leptogenesis,
in agreement with the data on neutrino masses and mixing. More
specifically, our post-inflationary setting favors the $A/H$
funnel and the $\tilde \chi^\pm_1-\chi$ coannihilation regions of
CMSSM with gravitino heavier than about $10~\TeV$. Leptogenesis is
realized through the out-of equilibrium decay of the inflaton to
the right-handed neutrinos $N_1^c$ and/or $N_2^c$, with masses
lower than $3.5\cdot10^{13}~\GeV$, and reheat temperature $\Trh$
close to $10^9~\GeV$.
%


\acknowledgments

C.P. acknowledges the Bartol Research Institute and the Department
of Physics and Astronomy of the University of Delaware for its
warm hospitality, during which this work has been initiated. He
also acknowledges useful discussions with G.~Lazarides and
S.~Martin. Q.S. acknowledges support by the DOE grant No.
DE-SC0013880.


\def\ijmp#1#2#3{{\sl Int. Jour. Mod. Phys.}
{\bf #1},~#3~(#2)}
\def\plb#1#2#3{{\sl Phys. Lett. B }{\bf #1}, #3 (#2)}
\def\prl#1#2#3{{\sl Phys. Rev. Lett.}
{\bf #1},~#3~(#2)}
\def\rmp#1#2#3{{Rev. Mod. Phys.}
{\bf #1},~#3~(#2)}
\def\prep#1#2#3{{\sl Phys. Rep. }{\bf #1}, #3 (#2)}
\def\prd#1#2#3{{\sl Phys. Rev. D }{\bf #1}, #3 (#2)}
\def\npb#1#2#3{{\sl Nucl. Phys. }{\bf B#1}, #3 (#2)}
\def\npps#1#2#3{{Nucl. Phys. B (Proc. Sup.)}
{\bf #1},~#3~(#2)}
\def\mpl#1#2#3{{Mod. Phys. Lett.}
{\bf #1},~#3~(#2)}
\def\jetp#1#2#3{{JETP Lett. }{\bf #1}, #3 (#2)}
\def\app#1#2#3{{Acta Phys. Polon.}
{\bf #1},~#3~(#2)}
\def\ptp#1#2#3{{Prog. Theor. Phys.}
{\bf #1},~#3~(#2)}
\def\n#1#2#3{{Nature }{\bf #1},~#3~(#2)}
\def\apj#1#2#3{{Astrophys. J.}
{\bf #1},~#3~(#2)}
\def\mnras#1#2#3{{MNRAS }{\bf #1},~#3~(#2)}
\def\grg#1#2#3{{Gen. Rel. Grav.}
{\bf #1},~#3~(#2)}
\def\s#1#2#3{{Science }{\bf #1},~#3~(#2)}
\def\ibid#1#2#3{{\it ibid. }{\bf #1},~#3~(#2)}
\def\cpc#1#2#3{{Comput. Phys. Commun.}
{\bf #1},~#3~(#2)}
\def\astp#1#2#3{{Astropart. Phys.}
{\bf #1},~#3~(#2)}
\def\epjc#1#2#3{{Eur. Phys. J. C}
{\bf #1},~#3~(#2)}
\def\jhep#1#2#3{{\sl J. High Energy Phys.}
{\bf #1}, #3 (#2)}
\newcommand\jcap[3]{{\sl J.\ Cosmol.\ Astropart.\ Phys.\ }{\bf #1}, #3 (#2)}
\newcommand\njp[3]{{\sl New.\ J.\ Phys.\ }{\bf #1}, #3 (#2)}
\def\prdn#1#2#3#4{{\sl Phys. Rev. D }{\bf #1}, no. #4, #3 (#2)}
\def\jcapn#1#2#3#4{{\sl J. Cosmol. Astropart.
Phys. }{\bf #1}, no. #4, #3 (#2)}
\def\epjcn#1#2#3#4{{\sl Eur. Phys. J. C }{\bf #1}, no. #4, #3 (#2)}

\end{document}